\documentclass[12pt,letterpaper]{article}
\pdfoutput=1
\usepackage{graphicx,array}
\usepackage{color}
\usepackage{latexsym}
\usepackage{amsthm}
\usepackage{amsmath}
\usepackage{amssymb}
\usepackage{hyperref}
\usepackage{bbold}
\usepackage{float}
\usepackage{subfig}

\def\simgt{\mathrel{\lower2.5pt\vbox{\lineskip=0pt\baselineskip=0pt
           \hbox{$>$}\hbox{$\sim$}}}}
\def\simlt{\mathrel{\lower2.5pt\vbox{\lineskip=0pt\baselineskip=0pt
           \hbox{$<$}\hbox{$\sim$}}}}

\newcommand{\be}{\begin{equation}}
\newcommand{\ee}{\end{equation}}
\newcommand{\bea}{\begin{eqnarray}}
\newcommand{\eea}{\end{eqnarray}}
\newcommand{\Eq}[1]{Eq.~(\ref{#1})}
\newcommand{\Eqs}[2]{Eqs.~(\ref{#1}) and (\ref{#2})}

\newcommand{\vev}[1]{\langle #1 \rangle}
\newcommand{\GeV}{\textrm{ GeV}}

\hypersetup{colorlinks,citecolor= blue,linkcolor= black}

\begin{document}

\begin{flushright}
UCB-PTH-14/35 \\
\end{flushright}

\hfill

\vspace{4cm}

\begin{center}
{\LARGE\bf

Multiverse Dark Matter: SUSY or Axions
}
\\ \vspace*{0.5cm}

\bigskip\vspace{1cm}{
{\large Francesco D'Eramo, Lawrence J. Hall and Duccio Pappadopulo}
} \\[7mm]
 {\it 

 Berkeley Center for Theoretical Physics, Department of Physics, \\
     and Theoretical Physics Group, Lawrence Berkeley National Laboratory, \\
     University of California, Berkeley, CA 94720, USA} \end{center}
\bigskip
\centerline{\large\bf Abstract}

\begin{quote} \small
The observed values of the cosmological constant {\it and} the abundance of Dark Matter (DM) can be successfully understood, using certain measures, by imposing the anthropic requirement that density perturbations go non-linear and virialize to form halos.   This requires a probability distribution favoring low amounts of DM, i.e. low values of the PQ scale $f$ for the QCD axion and low values of the superpartner mass scale $\tilde{m}$ for LSP thermal relics.  In theories with independent scanning of multiple DM components, there is a high probability for DM to be dominated by a single component.  For example, with independent scanning of $f$ and $\tilde{m}$, TeV-scale LSP DM and an axion solution to the strong CP problem are unlikely to  coexist.  With thermal LSP DM, the scheme allows an understanding of a Little SUSY Hierarchy with multi-TeV superpartners.  Alternatively, with axion DM, PQ breaking before (after) inflation leads to $f$ typically below (below) the projected range of the current ADMX experiment of $f = (3 - 30) \times 10^{11}$ GeV,    
providing strong motivation to develop experimental techniques for probing lower $f$.    

\end{quote}

\newpage

\newpage

\tableofcontents
\newpage

\section{Introduction and Overview}
\label{sec:intro}
Our universe evolved from an era of radiation domination to one of matter domination and is currently transitioning to one of cosmological constant domination.  Thus, over at least ten decades of temperature, $T$, the energy density of the universe involves two unknown parameters
\be
\rho \, = \, \Lambda + \rho_m + \rho_{rad} \, \sim \, \Lambda  + \xi T^3 + T^4
\label{eq:rho}
\ee
where, in the final form, numerical factors are omitted.  The matter abundance is parameterized by $\xi$, the ratio of the matter energy density to the photon number density $\xi = \rho_m/n_\gamma$, and has both baryon and DM components $\xi = \xi_D + \xi_B$.  The simplicity of this result should not mask the presence of two key theoretical problems
\begin{itemize}
\item {\it The Cosmological Constant}  \hspace{0.2in} Why is $\Lambda$ so small?
\item {\it Why Now?}  \hspace{0.2in}  Why are $\Lambda$ and $\rho_m$ comparable today; i.e., why is  $\Lambda \sim \xi T_0^3$?
\end{itemize} 

The cosmological constant problem has a simple anthropic solution  \cite{Weinberg:1987dv}: typical cosmological density perturbations do not become non-linear unless
\be
\Lambda \, < \, Q^3 \xi_D^4 \, f(\xi_D, \xi_B)
\label{eq:CCW}
\ee 
where $Q \sim 2 \times 10^{-5}$ is the magnitude of the primordial perturbations as they enter the horizon and $f=1$ for $\xi_D \gg \xi_B$. The absence of virialized matter halos leads to a dilute gas of inflating particles without any large scale structure or observers.  While it is convincing that (\ref{eq:CCW}) is a necessary requirement for observers, the expected multiverse probability distribution for $\Lambda$, $dP \propto d \Lambda$, leads to $\Lambda$ typically two to three orders of magnitude larger than observed \cite{Martel:1997vi}.  Furthermore, this condition does not directly address the Why Now problem; there could be an arbitrary long delay between the formation of halos and the appearance of observers.

The multiverse produced by eternal inflation is infinite; hence, when computing probabilities it is necessary to regulate divergences.  In the Causal Patch measure \cite{Bousso:2006ev}, which removes the divergence by looking at finite regions around geodesics,  the number of observers is diluted by inflation unless the time of $\Lambda$-domination occurs after the era at which observers occur, $t_\Lambda > t_{obs}$ \cite{Bousso:2007kq, Bousso:2010zi, Bousso:2010im}.  Since the dilution is exponential, this measure effectively forces the anthropic constraint
\be
\Lambda \, < \, \frac{1}{G \, t_{obs}^2}
\label{eq:CCB}
\ee 
where $G$ is Newton's constant.   Remarkably, this solves both the Cosmological Constant and the Why Now problems \cite{Bousso:2007kq, Bousso:2010zi, Bousso:2010im}:  the most probable observed universes are those close to the boundary (\ref{eq:CCB}), having $t_\Lambda \sim t_{obs}$.  Today we are observing the onset of $\Lambda$-domination,  so that the prediction of $\Lambda$ from (\ref{eq:CCB}) is more successful than from (\ref{eq:CCW}).  Hence, in this paper we assume a measure that leads to (\ref{eq:CCB}); currently the only known such measure is the Causal Patch measure.

A prediction for the cosmological constant, whether by (\ref{eq:CCW}) or (\ref{eq:CCB}), follows if the only scanning parameter in the multiverse is $\Lambda$, and is apparently destroyed if other relevant parameters, ($Q, \xi_D, \xi_B, G, t_{obs}$), scan.  To prevent runaways there must be additional environmental selection.   For example, requirements of galactic halo formation and cooling were included in multi-parameter scans in  \cite{Tegmark:2005dy, Bousso:2009ks}.  In a more general scanning of parameters, one hopes to discover that our universe lies at the tip of a multi-dimensional cone formed by anthropic boundaries from a variety of cosmological and particle physics requirements \cite{Hall:2007ja}.  For example, the virialization constraint (\ref{eq:CCW}) is a surface in this multi-dimensional space; a slice through the parameter space at fixed $(Q, \xi_D, \xi_B)$ leads to the original interpretation \cite{Weinberg:1987dv} of (\ref{eq:CCW}) as a bound on $\Lambda$. 

In general, the multiverse probability distribution depends on $n$ scanning parameters, and can be integrated over $m$ parameters to yield an effective distribution for the remaining $n-m$ parameters.  Differing subspaces are of interest for different problems.  For the evolution of our universe the subspace $(\xi_D, \xi_B, \Lambda)$ is key, since these are the parameters that enter (\ref{eq:rho}).  For simplicity, in this paper we set the baryon density to its observed value, $\xi_B = \xi_{B0}$, and scan in the 2d subspace $(\xi_D, \Lambda)$.  

We take a probability distribution
\be
dP \, = \, f(\xi_D, \Lambda) \; d \ln \xi_D \;  d \ln \Lambda,
\hspace{.5in} f(\xi_D, \Lambda) \, = \, p(\xi_D)  \Lambda \; n_{obs}(\xi_D, \Lambda)
\label{eq:P}
\ee
where $p(\xi_D) \, \Lambda$ is the a priori distribution, suitably marginalized over other scanning parameters, while $n_{obs}$ is the environmental weighting factor that we take to contain three contributions
\be
n_{obs}(\xi_D, \Lambda) \, = \, n_{meas}^\Lambda (\Lambda) \, n_{meas}^{\xi_D} (\xi_D) \, n_{vir}(\Lambda, \xi_D). 
\label{eq:nobs}
\ee
$n_{meas}^\Lambda$ contains the exponential dilution factor from the measure that imposes (\ref{eq:CCB}), while  
\be
n_{meas}^{\xi_D} (\xi_D) \, = \, \frac{\xi_B}{\xi_B + \xi_D}
\label{eq:nobsDB}
\ee
results from the dilution of baryonic observers for $\xi_D>\xi_B$ \cite{Freivogel:2008qc, Bousso:2013rda}  implied by such measures, in particular by the causal patch measure.  Finally, $n_{vir}(\Lambda, \xi_D)$ is the environmental weighting factor that follows from requiring density perturbations to go non-linear and virialize.
 
\begin{figure}[h!]
\begin{center} \includegraphics[width=0.6 \textwidth]{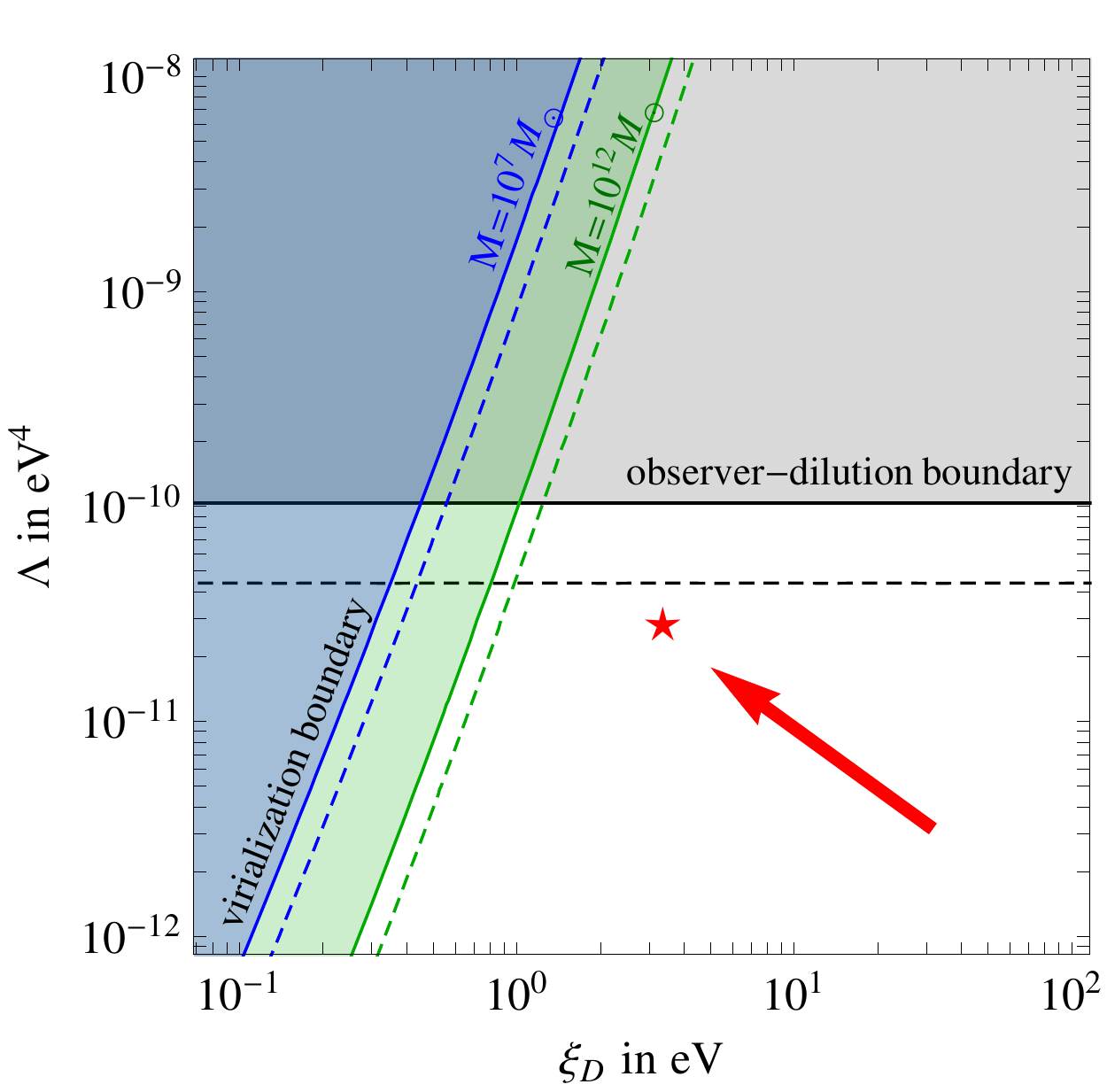}\end{center}
\caption{ The virialization (green: $M= 10^{12} M_\odot$; blue: $M= 10^7 M_\odot$) and observer-dilution boundaries in the ($\xi_D, \Lambda$) plane.   Shaded regions with solid line borders have exponential suppression of observers by at least 1\%, while the dashed lines are for 5\% suppression, as discussed in more detail in section \ref{sec:HV}.  A multiverse probability force, illustrated by the red arrow, allows our universe, the red star, to be typical.  The component of the force to large $\Lambda$ is predicted, while that to low $\xi_D$ is assumed.
\label{fig:Lambda-xi}    }
\end{figure}

In much of the $(\xi_D, \Lambda)$ plane the probability is exponentially suppressed; at low $\xi_D$ from the tail of the Gaussian distribution of the initial density perturbations, and at large $\Lambda$ from dilution of observers by inflation.  The onset of this exponential behavior can be approximated by catastrophic boundaries, and in Figure \ref{fig:Lambda-xi} we show the virialization and observer-dilution boundaries from the exponential behavior of  $n_{vir}(\Lambda, \xi_D)$ and $n_{meas}^\Lambda (\Lambda)$.  These boundaries correspond to (\ref{eq:CCW}) and (\ref{eq:CCB}).  The transition at these boundaries is quite sudden; as the shaded regions of  Figure \ref{fig:Lambda-xi} are entered, the probability of observers is exponentially reduced. 

On the other hand we assume that the a priori distribution $p(\xi_D)  \Lambda$ varies much more slowly and, in the region of interest, we take $p(\xi_D)$ to be a power law.   In the limit of taking sharp catastrophic boundaries, the probability distribution in the unshaded observer region is  $f(\xi_D, \Lambda) = p(\xi_D)  \Lambda \, n_{meas}^{\xi_D} (\xi_D)$.   If the probability force $(\partial \ln f / \partial \ln \xi_D, \partial \ln f / \partial \ln \Lambda)$ points towards the catastrophic boundaries, as shown by the red arrow in Figure \ref{fig:Lambda-xi}, typical observers should lie near the tip of the cone formed by the intersection of the two boundaries, as is the case for our universe shown by the red star.

At fixed $\xi_D = \xi_{D0}$, our universe is a factor of $10^2 (10^3)$ in $\Lambda$ from the most probable universes with halos of mass $M = 10^{12} (10^7) \, M_{\odot}$ \cite{Martel:1997vi, Bousso:2007kq}.  Similarly, from Figure \ref{fig:Lambda-xi} our universe is a factor of $10^2 (10^3)$ in $\Lambda$ from the virialization boundary for halos of mass $M = 10^{12} (10^7) \, M_{\odot}$.  However, as the virialization boundary is so steep, reflecting the fourth power of $\xi$ in (\ref{eq:CCW}),
at fixed $\Lambda = \Lambda_0$, our universe is only a factor 3-5 from the boundary in $\xi_D$.  
Hence, it is reasonable to take the view that $\Lambda$ is determined by observer dilution and $\xi_D$ by virialization.  In this 2d plane we simultaneously make use of (\ref{eq:CCW}) and (\ref{eq:CCB}) to understand both the size of the cosmological constant and the amount of DM
 \be
\Lambda \, \sim \, \frac{1}{G \, t_{obs}^2} \hspace{1in} \xi_D \, \sim \, \frac{\Lambda^{1/4}}{Q^{3/4}}.
\label{eq:obsdil+vir}
\ee 
We stress that the virialization boundary, used in \cite{Weinberg:1987dv} to predict a non-zero cosmological constant, is here used to determine the abundance of DM.  

We note that $\xi_D$ is the relevant variable to describe the abundance of DM; it is the quantity computed directly in any theory of DM genesis.  For example, models of WIMP DM characterized by a single mass scale $m$ lead to a freeze-out abundance $\xi_D \propto m^2$.  In this example, the DM abundance in our universe corresponds to a value for $m$ approximately a factor of 2 from the virialization boundary.

Other boundaries, in particular the requirement of halo cooling \cite{Tegmark:2005dy, Bousso:2009ks},  could play an important role in the $(\xi_D, \Lambda)$ plane.  However, the physics of this boundary is more complicated than for the virialization and observer-dilution boundaries, involving certain details of atomic and molecular physics, merger statistics and shocks, and has power dependence on the halo mass.   Hence we neglect halo cooling in this paper; inclusion would not significantly change our results.

In section \ref{sec:HV} we review the virialization boundary.  We stress that starting with our universe and decreasing $\xi_D$ leads rapidly to the loss of large scale structure.  In our view, virialization forces our universe to contain DM.  Of course, this view requires a distribution $p(\xi_D)$ {\it favoring low $\xi_D$} -- there is a cost in the multiverse to produce DM.  In the rest of the paper we explore the consequences of this important requirement.\footnote{In \cite{Bousso:2009ks} a probability distribution favoring low $\xi_D$ was crucial in understanding halo virialization and cooling time scales as well as the galactic mass scale.}  In section \ref{sec:xiddist} an effective distribution for $\xi_D$ is obtained by integrating over the cosmological constant.  For conventional LSP thermal freeze-out, the effective distribution for the superpartner mass scale $\tilde{m}$ should favor low values and, depending on the model, could lead to the TeV scale.  For QCD axion DM the effective distribution should favor a low PQ breaking scale, frequently leading to  $f \sim (10^{10} - 10^{11})$ GeV.

In section \ref{sec:singlecomp} we show that a probability distribution favoring low $\xi_D$ leads to a strong expectation that DM is dominated by a single component.\footnote{This is opposite to the case of a distribution favoring large $\xi_D$ and an anthropic boundary that excludes too much DM; if these boundaries are relevant DM could be multi-component \cite{Hall:2011jd}.   However, such proposed boundaries involve close stellar encounters or disk fragmentation, and appear less robust than the requirement that density perturbations go non-linear.   Another possibility is that the measure factor of eq. (\ref{eq:nobsDB}) causes the probability distribution to peak at $\xi_D \sim \xi_B$, explaining the observation that these matter components are broadly comparable \cite{Bousso:2013rda}; this scheme is consistent with either single- or multi-component DM.}    To be close to the virialization boundary we argue that all components of DM must have distributions favoring low values, and the dominant component will be the one with weakest distribution, with the others highly subdominant.  For a supersymmetric theory with both a thermal freeze-out LSP and an axion this would mean either $\tilde{m} \sim \text{TeV}$ and $ f \ll (10^{10} - 10^{11})$ GeV  or $\tilde{m} \ll \text{TeV}$ and $ f \sim (10^{10} - 10^{11})$ GeV; both are experimentally excluded.

LSP DM in supersymmetric theories with $\tilde{m}$ scanning is considered in more detail in section \ref{sec:LSP}.  In the case that weak-scale superpartners lead to too little DM, as would happen with a wino or Higgsino LSP, the scale of superpartners will necessarily be lifted to allow sufficient DM for virialization to occur, thereby generating a Little Susy Hierarchy with multi-TeV superpartners.   

In section \ref{sec:axion} we study QCD axion DM in more detail.  We begin by elucidating the parameter space of this cosmology, including both Pre- and Post-Inflation cases.  We determine which power law probability distributions for $f$ lead to the observed proximity of our universe to the virialization boundary, and obtain statistical predictions for the value of $f$ in our universe in the Pre-Inflation cosmology.  We find $f$ is frequently (always) {\it below} the expected reach of the ADMX experiment for Pre- (Post-) Inflation cosmologies.  New experimental techniques to probe this low $f$ window are urgently needed.  For our universe to be typical with a distribution favoring low $f$, thermal axion universes with $f \sim 10^{5\pm1}$ GeV must be catastrophic for observers.  We study thermal axion universes and conclude that observers are likely to be suppressed.


\section{Halo Virialization: the requirement of Dark Matter}
\label{sec:HV}

Halo Virialization on comoving scale $\lambda$ occurred when the matter density perturbation $\delta_m(\lambda)$ went non-linear.  Taking a wide view of the history of our universe this is a recent phenomenon.   Perturbations on the scale of our galaxy typically went non-linear at a redshift of a few.  As we look at other regions of the multiverse, we find that virialization is far from guaranteed.  The vast majority of the multiverse has values of the cosmological constant, $\Lambda$, far too large to allow relevant perturbations to go non-linear.  We begin by considering universes where the amount of DM varies but all other fundamental parameters, including $\Lambda$, are fixed at their observed values.   Later in this section we also scan $\Lambda$ and consider the virialization boundary in the $(\xi_D, \Lambda)$ plane.

$\xi_{B,D}$ are defined to be the ratio of the baryon and DM energy densities to the number density of photons
\be 
\xi_{B,D} \, = \, \frac{\rho_{B,D}}{n_\gamma}
\label{eq:xiBD}
\ee
and are determined by baryogenesis and DM genesis. Varying $\xi_D$ induces a variation in $T_{e, \Lambda}$, the temperatures of matter-radiation and matter-$\Lambda$ equality, as well as the time-temperature relation.\footnote{The quantities $\Omega_{B,D}$ are awkward as they depend on the critical density, which must be carefully defined as parameters are scanned; nevertheless, ratios are simple: $\Omega_B/\Omega_D = \xi_B/\xi_D$ and $\Omega_D/\Omega_{D0} = \xi_D/\xi_{D0}$.}  Ignoring all numerical factors, the parametric form for the energy density is $\rho \sim T^4 + \xi T^3 + \Lambda$, with terms for radiation, matter and the cosmological constant, and $\xi = \xi_B + \xi_D$.  Hence $T_e \sim \xi$ and $T_\Lambda \sim (\Lambda/\xi)^{1/3}$.

\begin{figure}
\begin{center}
\includegraphics[scale=0.67]{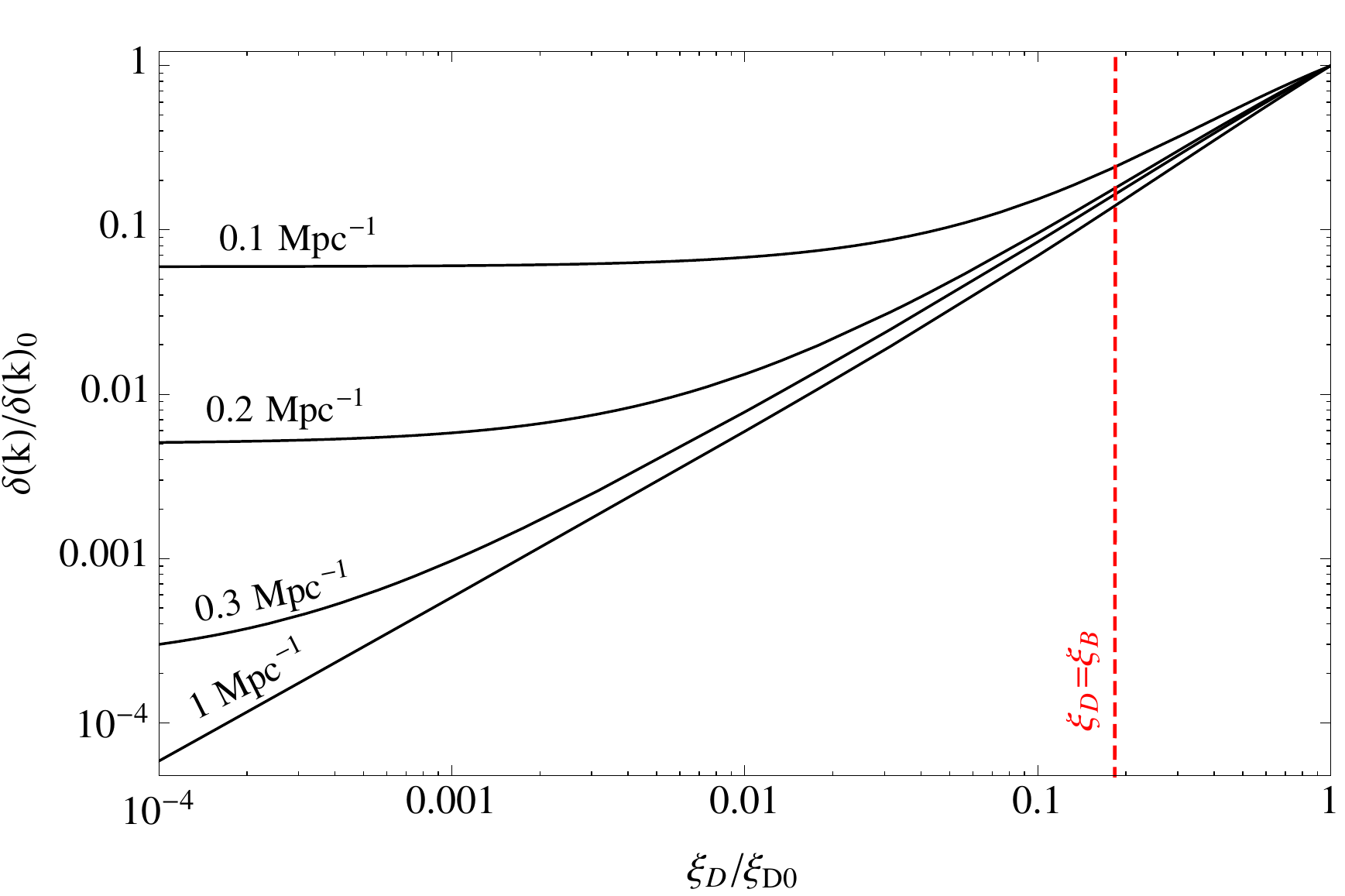}
\end{center}
\caption{Damping of matter density perturbations as the DM abundance is reduced, with fixed $\Lambda = \Lambda_0$ and $\xi_B = \xi_{B0}$.}
\label{fig:DarkDominance}
\end{figure}

The evolution of density perturbations is well-known and has been studied at a high level of accuracy.  In Figure \ref{fig:DarkDominance}, using a fitting formula of \cite{Eisenstein:1997ik}, we plot the ratio $\delta_m/\delta_{m0}$, at an era that is well after recombination and matter-radiation equality but still in the perturbative linear regime, against $\xi_D/ \xi_{D0}$.  (The subscript ``0" refers to values in our universe.) The four curves are for comoving scales of $k = (0.1 - 1) \mbox{Mpc}^{-1}$, corresponding to baryonic masses of $M \sim (10^{15} - 10^{12}) M_\odot$, where $M_\odot$ is the mass of our sun.  The most striking feature is that a reduction in the amount of DM causes a very substantial drop in the density perturbation and, for masses of our galaxy and smaller, this reduction becomes extreme at very low DM abundances.

This behavior can be understood from the rough analytical approximation
 \be
\delta_m(M) \, \sim \, \frac{T_e}{T_\Lambda} \left( f_B \; e^{-(\frac{M_S}{M})^{0.47}} \: + \; f_D  \, G_{rad}(M,T_e) \right) \, Q_0
\label{eq:deltam}
\ee
where $f_{B,D} = \xi_{B,D} /(\xi_D + \xi_B)$ are the fractions of matter contained in baryon and dark species, and the density perturbations are taken to enter the horizon with size $Q_0 = 2 \times 10^{-5}$, independent of their mass $M$.  During the matter dominated era, there is a linear growth in the perturbations resulting approximately in the overall factor 
\be
\frac{T_e}{T_\Lambda} \sim \left( \frac{\xi^4}{\Lambda} \right)^{1/3}
\label{eq:TeTL}
\ee
(which assumes the linear regime).  The two terms in (\ref{eq:deltam}), proportional to  $f_{B}$  and $f_{D}$, originate from baryon and DM perturbations and evolve very differently.  On mass scales less than $M_S$ the baryon perturbations are reduced by photon diffusion (Silk) damping near the era of recombination, $M_{S0} \sim 10^{16} M_\odot$.  On the other hand the DM is not coupled to the photons and is able to grow logarithmically during the radiation dominated era by a factor $G_{rad}$.   After recombination, the baryons are released from the coupling to photons and the perturbations of baryons and DM can be combined into a single matter perturbation, $\delta_m$ of Eq.~(\ref{eq:deltam}).   

In our universe the baryon perturbations on scales relevant for virialization are strongly Silk damped, but this is almost irrelevant as after recombination the baryons fall into the potential of the DM perturbations, which are large since $f_D G_{rad} \gg f_B$.    Returning to Figure \ref{fig:DarkDominance}, the lower curve is for halo mass $M \sim 10^{12} M_\odot$, sufficiently low that the baryon contribution to the perturbation is negligible and the resulting perturbation drops linearly with $f_D$.  The curves at successively larger $M$ eventually reach a plateau where the baryon contribution, even though damped, dominates the dark perturbation at low $f_D$.

As $\xi_D$ is reduced below $\xi_{D0}$ the reduction in $\delta_m$ is immediate.  Even though $f_D$ at first stays close to unity, the matter dominated growth factor $T_e/T_\Lambda$ scales as $\xi_D^{4/3}$.  Once the DM density drops below that of baryons, $T_e/T_\Lambda$ becomes constant, but $f_D$ scales as $\xi_D$ and drops below unity.  The second term of (\ref{eq:deltam}) is then reduced because, when the baryons are released from the photon coupling after recombination, there is only a small potential from the DM for them to fall into.  The numerical solution shown in Figure \ref{fig:DarkDominance} is smooth at $\xi_D = \xi_{B0}$, but asymptotically the slope varies from 4/3 to 1, for perturbations where the baryon component is Silk damped.

From Figure \ref{fig:DarkDominance} we see that when $\xi_D/\xi_{D0} \sim 1/5$ the matter perturbation on galactic scales and smaller is already suppressed by an order of magnitude.  This means that typical perturbations on these scales never go non-linear as their growth is halted by the onset of $\Lambda$ domination.  On rare occasions the perturbations may fluctuate to larger values, allowing regions of growth, but this becomes exponentially rare as $\xi_D$ drops further.

We obtain a virialization boundary, shown in Figure \ref{fig:Lambda-xi}, by fitting the curves in Figure 2 of \cite{Bousso:2007kq} by an exponential. With good accuracy the $n_{vir}(\Lambda, \xi_D)$ factor depends only on the halo mass and the value of the matter density perturbation $\delta_m(t_\Lambda)$ at the time of cosmological constant domination. 
We consider two values for the halo mass $M=10^{(7,12)}M_{\odot}$ and find
\bea\label{eq:boundaryvir}
n_{vir}(\delta_m, 10^7M_{\odot})&\propto& e^{-0.03 (\delta_{m0}/\delta_m)^{1.8}} \\ \nonumber
n_{vir}(\delta_m, 10^{12}M_{\odot})&\propto& e^{-0.14 (\delta_{m0}/\delta_m)^{1.9}}.
\eea
The solid (dashed) lines in Figure \ref{fig:Lambda-xi}  correspond to those points in the $(\xi_D, \Lambda)$ plane where the exponential factor in (\ref{eq:boundaryvir}) drops to 1\% (5\%).   Thus the excluded regions contain 1\% (5\%) of observers.  
For example, for $\Lambda = \Lambda_0$ and $M= 10^{(7,12)} M_\odot$ we find the 1\% boundary to be $\xi_{1\%} = (0.10, 0.23) \xi_0$ and the 5\% boundary to be $\xi_{5\%} = (0.12, 0.27) \xi_0$.

Figure \ref{fig:Lambda-xi} also shows the observer dilution boundary.  We define this by extracting the exponential behavior of $n_{meas}^\Lambda (\Lambda)$ from \cite{Bousso:2010zi}:
\be
n_{meas}^\Lambda (\Lambda)\propto e^{- 3(\Lambda/\Lambda_{obs})^{1/2}}. 
\label{eq:nLambda}
\ee
$\Lambda_{obs}$ is related to the time $t_{obs}$ at which observers occur by
\be\label{eq:Lambdaobsobs}
\Lambda_{obs}\equiv\frac{1}{G\, t_{obs}^2}\approx (2.6\times 10^3\, {\textrm{eV}})^4.
\ee
The solid (dashed) horizontal line in Figure \ref{fig:Lambda-xi} corresponds to a value of the exponential factor in (\ref{eq:nLambda}) of 1\% (5\%). 

Baryonic perturbations on scales larger than $M_S$ are not subject to Silk damping, and therefore could be relevant for large scale structure even if $\xi_D = 0$.    If such perturbations go non-linear it is possible that a top-down scheme for large scale structure could lead to observers.  The difficulty is estimating the probability of such observers relative to ones in universes similar to our own.  Since we assume a probability distribution for $\xi_D$ that decreases as $\xi_D$ drops below $\xi_{D0}$, there will be a gain in probability from reducing $\xi_D$ well below $\xi_{B0}$.  However this can be offset by the reduction in $\Lambda$ necessary to allow such large baryonic perturbations to go non-linear.  With $\xi_D$ irrelevantly small, $T_e$ drops by about a factor 6 from $T_{e0}$, and such large baryonic perturbations don't grow significantly until the matter dominated era is reached.  We estimate that such perturbations go non-linear only if $T_\Lambda$ is reduced by $\sim 10^2$ relative to our universe and the reduction in $\Lambda$ to achieve this costs $\sim 10^{-6}$ in probability.  Halo and star formation is entirely different in a top-down scheme, so that the corresponding observer weighting factors will be very different and are unknown.




\section{An Effective Distribution for $\xi_D$}
\label{sec:xiddist}

For the rest of the paper it is convenient to perform an approximate integral of the probability distribution over $\Lambda$ so that we can focus our discussion on the remaining effective probability distribution for $\xi_D$.  We define the observer region to be where the exponentials in $n_{vir}(10^{12}M_\odot)$ and $n_{meas}^\Lambda$, shown in (\ref{eq:boundaryvir}) and (\ref{eq:nLambda}), are larger than $\sim 50\%$.  Thus the observer region is similar but slightly smaller than the unshaded region in Figure 1, where the boundaries were defined by these exponentials reaching 1\%.  We define the observer dilution and virialization boundaries of the observer region to be 
\be
\Lambda < \Lambda_c, \hspace{0.5in} \xi_D > \xi_c \left( \frac{\Lambda}{\Lambda_c} \right)^{1/4}
\label{eq:obsbound}
\ee
so that they intersect at $(\xi_c, \Lambda_c) \sim (0.5\, \xi_{D0}, \Lambda_0)$.
Approximating $n_{vir}$ and $n_{meas}^\Lambda$ as $\theta$ functions at the boundaries (\ref{eq:obsbound}), in the observer region the probability distribution of (\ref{eq:P}, \ref{eq:nobs}, \ref{eq:nobsDB}) becomes
\be
dP \, = \, p(\xi_D) \, \frac{\xi_{B0}}{\xi_{B0} + \xi_D} \, d \ln \xi_D \, d \Lambda.
\label{eq:Pinobs0}
\ee
Here we use the value of $\xi_B$  observed in our universe as we are not scanning over the baryon density.

Integrating over $\Lambda$ gives an effective distribution for the single parameter $\xi_D$
\be
dP \, = \, p(\xi_D) \, \frac{\xi_{B0}}{\xi_{B0} + \xi_D} \, d \ln \xi_D \, 
\begin{cases}
                        \Lambda_c  & \xi_D > \xi_c \\
                        \Lambda_c (\xi_D/\xi_c)^4 & \xi_D < \xi_c
                    \end{cases}
\label{eq:Pinobs}
\ee
where the last factor arises because the maximum value of $\Lambda$ in the observer region depends on $\xi_D$.  We assume that the $(\xi_D/\xi_c)^4$ factor makes the region $\xi_D < \xi_c$ sufficiently improbable that there is little error in taking the 1d observer region to be
\be
\xi_D > \xi_c.
\label{eq:xic}
\ee
This assumption would only fail if the distribution $p(\xi_D)$ favored low $\xi_D$ so strongly that it overcompensates the $(\xi_D/\xi_c)^4$ factor.  However, this would cause runaway along the virialization boundary in Figure 1 to both low $\xi_D$ and low $\Lambda$, destroying the successful understanding of the Why Now problem.  For less extreme $p(\xi_D)$ distributions, the Why Now problem is solved, since the typical value of $\Lambda$ is $\Lambda_c \sim \Lambda_0$. In the rest of the paper we fix $\Lambda = \Lambda_0$ and perform a 1d scan over $\xi_D$ with distribution (\ref{eq:Pinobs}) subject to (\ref{eq:xic}).

\section{Single Component Dark Matter}
\label{sec:singlecomp}

\begin{figure}
\begin{center}
\includegraphics[height=2.9in]{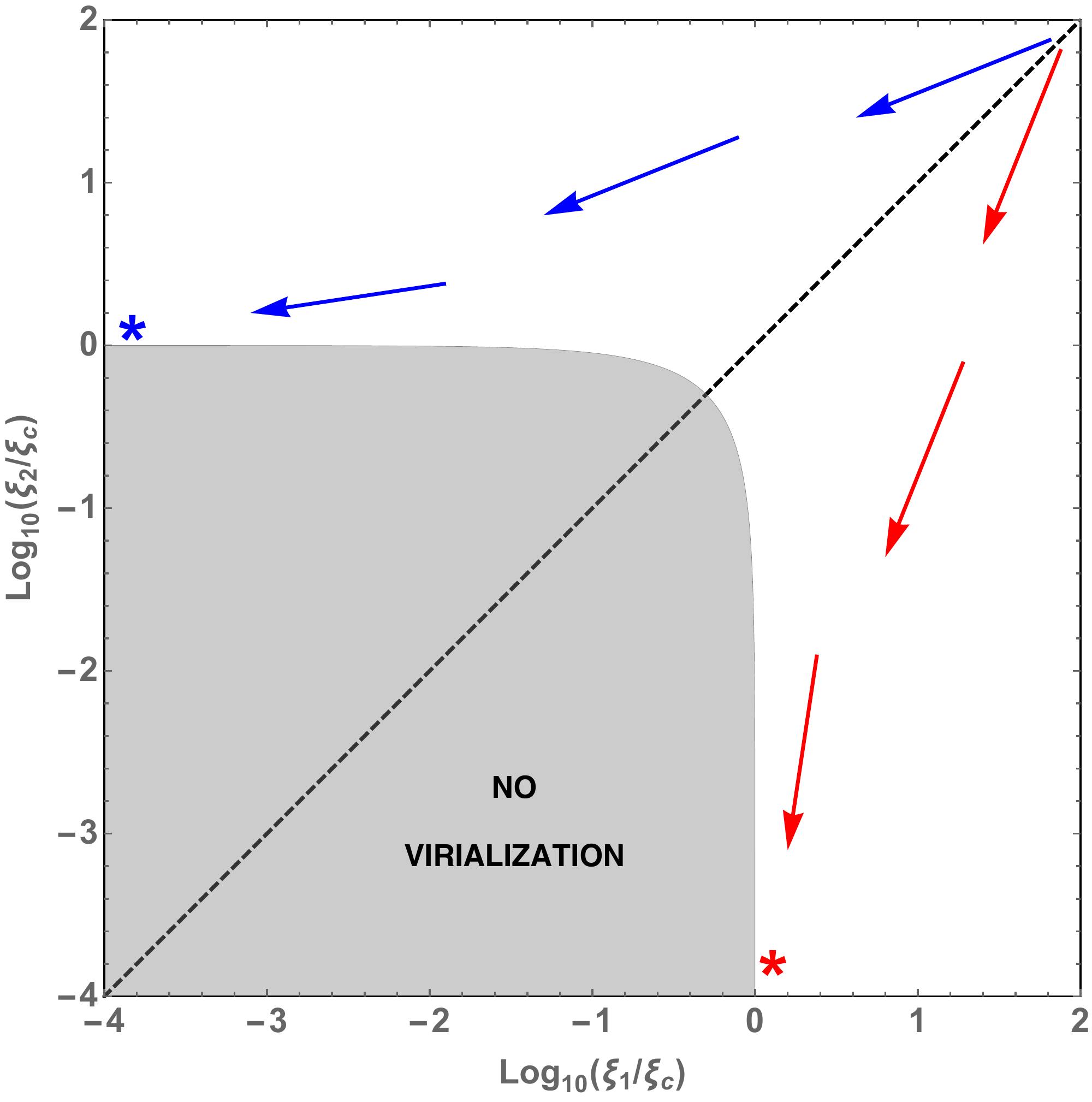}
\end{center}
\caption{Typical runaway behavior at the virialization boundary for two independently scanning contributions to DM, $\xi_1$ and $\xi_2$.  Virialization fails in the shaded region, where $\xi_1 + \xi_2 < \xi_c$.  The blue arrows represent one choice for $\nabla P_i$ which leads to runaway along the virialization boundary to small $\xi_1$, as shown by the blue star, and the red arrows represent another choice that leads to runaway to small $\xi_2$ and the red star.  
}
\label{fig:SingleComponent}
\end{figure}

The arguments of previous sections are independent of the nature of DM.  In this section we show that if the abundance of DM is explained by the virialization boundary it is dominantly composed of a single component.   Suppose that there are many components, $\xi_D = \Sigma_i \xi_i$, with sufficiently many scanning parameters that the energy densities, $\xi_i$, scan independently with prior $dP = \Pi_i \; P_i(\xi_i) \, d \ln \xi_i$.    In our universe, if some species has $\xi_i \gg \xi_c$ density perturbations are easily able to go non-linear, and we are far from the virialization boundary.  Hence, for the abundance of DM to be explained by proximity to the virialization boundary, all $\xi_i$ should be of order $\xi_c$ or smaller.  Is it possible that more than one component has $\xi_i \sim \xi_c$, with distributions favoring low $\xi_i$?  In general this possibility is extremely unlikely.  The component with the weakest distribution towards the boundary will have an abundance close to $\xi_c$, while the other components will typically have much smaller abundances, as illustrated in Figure \ref{fig:SingleComponent}.   Hence, if the virialization boundary is the correct explanation for the DM density, {\it DM is strongly dominated by a single component}.\footnote{This is very different from an anthropic boundary that places an upper limit on $\xi_D = \Sigma_i \xi_i$, such as close stellar encounters.  In this case, if several species have distributions that favor large values of $\xi_i$ then these components are all expected to have densities close to the boundary value \cite{Hall:2011jd}.}

How small might we expect the sub-dominant components to be? In generic theories of DM, $\xi_i$ can vary over many orders of magnitude.  A sub-dominant species must have the gradient $d P_i/ d \xi_i$ negative at $\xi_i = \xi_c$ so that $\xi_i$ will runaway to low some low value $\xi_{{\rm min} \, i}$ where the sign of $d P_i/ d \xi_i$ changes.  It would be accidental for this to happen anywhere near $\xi_c$, which is determined by the physics of virialization and is independent of the prior distribution.  Hence, we expect $\xi_{{\rm min} \, i}$ to be less than $\xi_c$ by at least a few orders of magnitude, as illustrated by the stars in Figure \ref{fig:SingleComponent}. 

How improbable is multi-component DM?  Consider a two-component model with densities $\xi_{1,2}$ scanning independently with an effective distribution
\be
\frac{d P}{d \log \xi_1 \; d \log \xi_2} \, = \, C \,\,  \theta\left( \xi_1 + \xi_2 - \xi_c \right) \frac{1}{1 + \frac{\xi_1 + \xi_2}{\xi_B}} \; \xi_1^{n_1} \, \xi_2^{n_2} 
\label{eq:TwoComponentP}
\ee
with  $n_{1,2} < 0$ and $C$ a normalization constant.  We take (\ref{eq:TwoComponentP}) to be valid over a wide range of $\xi_{1,2}$, including the region of the virialization boundary near $\xi_1 \sim \xi_2$, and down to near $\xi_{{\rm min} \, 1,2}$, where it breaks down and the gradient of the distribution changes sign.   Let $P_{\text{multi}}$ be the probability for $\xi_{1,2} \sim \xi_c$ integrated over a region with $\Delta \xi_{1,2} \sim \xi_{1,2}$.  Similarly let $P_{\text{single}\, 1}$ be the probability for $(\xi_1, \xi_2) \sim (\xi_c, \xi_{{\rm min} \, 2}) $ integrated over a region with $\Delta \xi_{1,2} \sim \xi_{1,2}$.  We find
\be
\frac{P_{\text{multi}}}{P_{\text{single}\, 1}} \, \sim \, \left( \frac{\xi_{{\rm min} \, 2}}{\xi_c} \right)^{|n_2|}.
\ee
In going from single component DM with $(\xi_1, \xi_2) \sim (\xi_c, \xi_{{\rm min} \, 2})$ to multi-component DM with $(\xi_1, \xi_2) \sim (\xi_c, \xi_c) $ one loses in probability from the $\xi_2^{n_2}$ factor of (\ref{eq:TwoComponentP}) by changing $\xi_2$ from its minimal value to $\xi_c$.   The result for $P_{\text{multi}}/P_{\text{single} \, 2}$ is obtained by interchanging $1 \leftrightarrow 2$.  For example, for $|n_i|=N$ and $\xi_{{\rm min} \, i} = 10^{-3} \xi_c$, multi-component DM is less probable than single component by a factor of $10^{3N}$.

Consider applying these general arguments to supersymmetric theories that contain a QCD axion, and assume a cosmology that leads to both misalignment axion DM, $\xi_a$, and thermal LSP freezeout DM, $\xi_{LSP}$.  Further assume that the multiverse distribution functions for the axion decay constant, $f$, and the scale of supersymmetry breaking, $\tilde{m}$, favor low values, so that the size of $\xi_D = \xi_a + \xi_{LSP}$ will be explained by proximity to the virialization boundary (since $\xi_a \propto f$ and $\xi_{LSP} \propto \tilde{m}^2$).  The relative strengths of the distributions for $f$ and $\tilde{m}$ are unknown, so we do not know which is the dominant component with density near $\xi_c$ and which the sub-dominant component with density at least a few orders of magnitude below $\xi_c$.   If axions are dominant $f \sim 10^{11}$ GeV, while if LSPs are dominant $\tilde{m} \sim $ TeV, (these are rough order of magnitude estimates, which have theoretical uncertainties and model dependencies, but are sufficient for our purposes).   This then implies that if axions are dominant $\tilde{m}$ is at least a few orders of magnitude below the TeV scale, while if LSPs are dominant $f$ is at least a few orders of magnitude below $10^{11}$ GeV.  Both cases are observationally excluded -- they do not describe our universe.  Hence, in our scheme, with the abundance of DM explained by closeness to the virialization boundary, conventional TeV-scale LSP freezeout DM is {\it inconsistent} with the axion solution to the strong CP problem.   To avoid this conclusion, $\xi_{{\rm min} \, a, LSP}$ must {\it both} be accidentally close to $\xi_c$, implying that the observed DM abundance is close to the peak of the 2d probability distribution. 

Of course, theories can contain either stable TeV-scale LSPs or axions, and these both remain interesting DM candidates when the abundance is explained by the virialization boundary.   We consider the dominant DM component to be the lightest superpartner in section \ref{sec:LSP} and the axion in section  \ref{sec:axion}.

\section{LSP Dark Matter and a Little SUSY Hierarchy}
\label{sec:LSP}

In the Standard Model the weak scale, $v$, is fine-tuned.  Without a multiverse this suggests that some new physics, such as supersymmetry, should occur at the weak scale.  However, in a multiverse such new physics is not required, since a finely-tuned weak scale may result from anthropic requirements \cite{Agrawal:1997gf, Hall:2014dfa}.  Whether new physics is likely at the weak scale depends on probability distributions.  

Consider a supersymmetric extension of the Standard Model with the overall mass scale of the superpartners scanning, but with fixed superpartner mass ratios.  For convenience we define $\tilde{m}$ to be the mass of the Lightest SuperPartner (LSP), which we assume is cosmologically stable and has a thermal freeze-out abundance $\xi_D \propto \tilde{m}^2$. Using  (\ref{eq:Pinobs}, \ref{eq:xic}) and assuming a prior distribution $\tilde{m}^n$, we obtain an effective multiverse probability distribution
 \be
dP \, \propto \, \theta(\tilde{m} - \tilde{m}_c) \; \tilde{m}^n \left(\frac{1}{1 + 6.0(\tilde{m}/\tilde{m}_0)^2}\right) 
\left(\frac{v^2}{v^2 + \tilde{m}^2}\right) \, d \ln \tilde{m}.
\label{eq:mtildedist}
\ee
The first factor in parenthesis is the measure factor of (\ref{eq:nobsDB}), with $\tilde{m}_0$ the value of the LSP mass that leads to $\xi_D = \xi_{D0}$.    The last factor in parenthesis arises from fine-tuning, to satisfy the anthropic electroweak symmetry breaking requirement, and for $\tilde{m} \gg v$ becomes  $v^2/\tilde{m}^2$.  It is an additional factor in the environmental weighting quantity $n_{obs}$, omitted in (\ref{eq:nobs}) since it is special to LSP DM.  For simplicity we have assumed that the superpartners relevant for this tuning have mass close to $\tilde{m}$.  As discussed in section 3, the argument of the theta function that approximates the virialization boundary corresponds to the weighting factor $n_{vir} = 0.5$, leading to $\tilde{m}_c \simeq 0.7 \tilde{m}_0$.

A crucial question is the size of $\tilde{m}_0$, which then determines $\tilde{m}_c$.  So far our argument is quite general, with little dependence on the details of the supersymmetric model.   However, $\tilde{m}_0$ is model dependent, depending on the nature of the LSP and its interactions.  As is well-known, freeze-out frequently involves a mass scale somewhat larger than the weak scale.  For example, $\tilde{m}_0 \simeq 1 (3)$ TeV for pure Higgsino (wino) DM.   We assume $\tilde{m}_0$ is larger than the weak scale and, for illustration, we take $\tilde{m}_0 = 2.2$ TeV, which gives $\tilde{m}_c = 1.5$ TeV.

 \begin{figure}
\begin{center}
\includegraphics[scale=0.53]{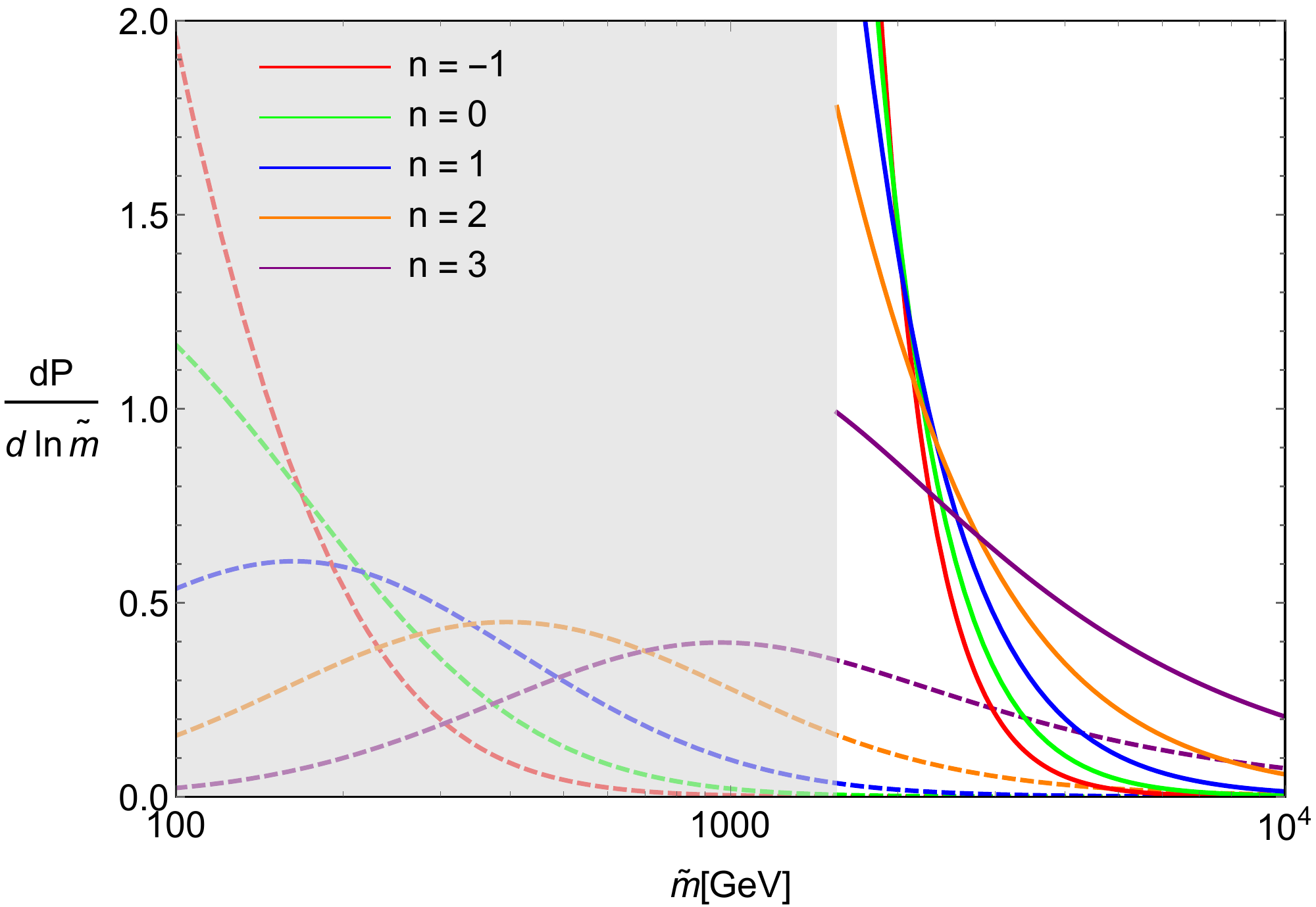}
\end{center}
\caption{The probability distribution (\ref{eq:mtildedist}) for the LSP mass for various values of $n$ (describing the prior distribution, $\tilde{m}^n$) and for $\tilde{m}_0 = 2.2$ TeV.   Solid (dashed) curves include (ignore) the theta function, $\theta(\tilde{m} - \tilde{m}_c)$, that approximates the virialization boundary. 
Imposing the requirement of virialization, which excludes the shaded region and changes the distributions from the dashed to solid curves,  leads to a Little Susy Hierarchy for a wide range of prior distributions, $n<4$. }
\label{fig:LittleSUSYhierarchy}
\end{figure}

In Figure \ref{fig:LittleSUSYhierarchy} the dashed curves show the probability distribution (\ref{eq:mtildedist}) for various $n$, with the $\theta$ function representing the virialization boundary ignored.   For $n<0$ the distributions are peaked at low $\tilde{m}$.   For $0<n<2$ they are peaked near $v$, so that superpartners are expected at the weak scale.   For $2<n<4$ they are peaked near $\tilde{m}_0/\sqrt{6} \sim 900$ GeV, as illustrated by the dashed purple line.  It is in this range of $n$ that the multiverse predicts comparable amounts of dark and baryonic matter since the peak in the distribution is determined by the $1/(1 + \xi_D/\xi_B)$ factor \cite{Bousso:2013rda}.   However, as shown by the shading in Figure \ref{fig:LittleSUSYhierarchy}, these distributions are all peaked in the region forbidden by the virialization requirement.  Including the theta function in the distribution (\ref{eq:mtildedist}) to enforce virialization, the distributions shift from the dashed curves to the solid curves.  For all $n<4$ these are peaked at the virialization boundary, leading to the expectation of a Little Supersymmetric Hierarchy with superpartners in the TeV domain.    Of course the detailed superpartner spectrum depends on the model, and may be compressed or split to varying degrees.   We stress that a wide range of prior distributions, even those that strongly favor large values of supersymmetry breaking, lead to proximity to the virialization boundary and to a Little Hierarchy.\footnote{The fine-tuning and measure factors both serve to prevent supersymmetry breaking running away to high values.  However, very strong prior distributions with $n>4$ beat these factors, so the LSP is typically many orders of magnitude heavier than the weak scale and is likely irrelevant for DM.}
 
For $n<0$ most universes have $\tilde{m} \ll v$ and in this region $\xi_D$ may have a different dependence on $\tilde{m}$.  For example, if LSP annihilation occurs via virtual weak gauge bosons the annihilation cross section scales as $\tilde{m}^2/v^4$ so that  $\xi_D \propto \tilde{m}^{-2}$, giving sufficient DM to allow large scale structure formation at $\tilde{m} \ll v$. However, in many supersymmetric theories, such as the MSSM, it is not possible (or requires a costly fine-tune) to obtain $v \gg \tilde{m}$, so these universes are excluded from anthropic requirements on the size of the weak scale.  Hence, in such supersymmetric theories, with $n<0$, one again has typical observers with $\tilde{m}$ just above $\tilde{m}_c$.
 


\section{Axion Dark Matter}
\label{sec:axion}

The QCD axion  \cite{Weinberg:1977ma, Wilczek:1977pj} provides a minimal extension of the Standard Model (SM) that provides both a solution of the strong CP problem \cite{Peccei:1977hh} and cosmological cold DM  \cite{Preskill:1982cy, Abbott:1982af, Dine:1982ah}.  Over a period of several decades the importance of a QCD axion has steadily grown due to the absence of any other plausible solution to the strong CP problem, together with a succession of null results searching for weak-scale DM from both galactic DM and particle accelerators.   The axion is a pseudo-Goldstone boson produced at scale $f$ by the spontaneous breaking of a Peccei-Quinn (PQ) symmetry that carries a QCD anomaly.

We fix our notation by introducing the effective lagrangian we will use to discuss axion physics.  The PQ symmetry is broken dominantly by a canonically normalized scalar field with vacuum structure $\vev{\phi} = V e^{ia(x)/V}$, where $a(x)$ is the axion field. 
At energies below the PQ breaking scale and the heavy SM fermion masses, but above the QCD confinement scale, the most general axion interactions are described by the following effective lagrangian at order $1/f$
\be \label{effaxion}
\mathcal L_a=\frac{1}{2}\, \partial^\mu a \, \partial_\mu a-\frac{a}{f}\frac{\alpha_3}{8\pi} \, G^{a\mu\nu}\tilde G_{\mu\nu}^a-c_{\gamma}\frac{a}{f}\frac{\alpha}{8\pi} \, F^{\mu\nu}\tilde F_{\mu\nu}+  \frac{\partial_\mu a}{f} \sum_i \left(c^V_i\bar \psi_i \gamma^\mu \psi_i+c^A_i \bar \psi_i \gamma^\mu\gamma_5 \psi_i \right).
\ee
The sum is over the light SM fermion fields $\psi_i= u, d, e, \nu$. Notice that (\ref{effaxion}) defines the constant $f$ through the normalization of the $G\tilde G$ term; it is related to the vev $V$ by $V = Nf$, where $N$ is an integer determined by the color anomaly of the PQ symmetry, and is known as the domain wall number.  

QCD instantons and the PQ color anomaly generate a small zero-temperature axion potential $\approx \Lambda_{QCD}^4 ( 1 - \cos a/f)$, leading to $N$ equivalent vacua, and an axion mass that can be computed reliably using chiral perturbation theory 
\be\label{mazero}
m_a(T=0)=\frac{\sqrt{m_u/m_d}}{1+m_u/m_d}\frac{f_\pi m_\pi}{f}= 6\times 10^{-6}\, {\textrm{eV}}\left(\frac{10^{12}\,{\textrm{GeV}}}{f}\right).
\ee

\subsection{The Misalignment Mechanism and the Virialization Boundary}
\label{subsec:axion}

Since all axion interactions are suppressed by the scale $f$, for large values of $f$ the axion is decoupled from the thermal bath in the early universe.
The behavior of the classical axion field during these epochs is characterized by two phases. For $H\gg m_a$, the axion field is stuck at a random position in field space $a(x)=a_i$. As soon as $H\ll m_a$ the Hubble friction becomes irrelevant and the axion field starts to behave as in flat space oscillating around its equilibrium position, $a=0$. Once the anharmonic corrections to the potential are small the energy density stored in these oscillations dilutes as non-relativistic matter with the Hubble flow and contributes as a cold component of the DM. This mechanism, which generates a contribution to the cosmological cold DM density through the coherent oscillations of a classical scalar field, is called the ``misalignment mechanism''.  


In the ``Pre-Inflation" cosmology, where the PQ phase transition occurs before or during inflation, initial fluctuations of the axion field on scales corresponding to our Hubble horizon can be neglected, so that the initial misalignment angle $\theta=a_i/f$ is a constant.  The axion DM energy density from misalignment has been computed to be \cite{Bae:2008ue,Kim:2008hd}
\be\label{rhoaxion}
\rho_{mis}^{\text{Pre}}(f, \theta) \, \approx \, (2.0 \times 10^{-3}\,{\textrm{eV}})^4\left(\frac{f}{10^{12}\,{\textrm{GeV}}}\right)^m\theta^2 F(\theta) 
\ee
with $m \simeq 1.2$.  This result applies for $f\lesssim 10^{15\div 16}$\,GeV, so that the axion field starts to oscillate at a temperature above $\Lambda_{QCD}$, and includes all values of $f$ of interest to us.  The function $F(\theta)$ corrects for the anharmonicities of the potential, and has normalization $F(0)=1$.   This result has theoretical uncertainties arising from a non-perturbative instanton calculation of the temperature dependent axion mass \cite{Wantz:2009mi}.  Hence we introduce a corresponding uncertainty in the value of $f$ that accounts for DM in the Pre-Inflation cosmology, and for illustration adopt the range
\be 
f^{\text{Pre}}(\theta)  \, = \, \frac{1}{(\theta^2  F(\theta))^{0.8}} \; \; \, (3 - 10) \times 10^{11} \, \text{GeV} .
\label{eq:fPre}
\ee

Since misalignment axion production occurs near the QCD scale, the DM abundance of (\ref{rhoaxion})  is quite robust to non-standard cosmologies at higher temperatures.  However, the abundance could be affected by changing the cosmology below the QCD scale; for example, entropy production would require larger values of $f$ to explain the observed DM abundance.

In the ``Post-Inflation" cosmology, where the PQ phase transition occurs after inflation, (\ref{rhoaxion}) has to be averaged over the possible values of $\theta$ 
\be
\langle \theta^2 F(\theta) \rangle=(2\pi)^{-1}\int_{-\pi}^{\pi} d\theta\, \theta^2 F(\theta),
\ee
where the initial misalignment angle is assumed to be uniformly distributed between -$\pi$ and $\pi$, yielding
\be \label{rhoaxion2}
\rho_{mis}^{\text{Post}}(f) \, \approx \, (3.0 \times 10^{-3}\,{\textrm{eV}})^4\left(\frac{f}{10^{12}\,{\textrm{GeV}}}\right)^m.
\ee
If this were the only source of axion DM, the required value of $f$ would be centered on $1.5 \times 10^{11}$ GeV.  However, a network of axion strings is formed at the PQ phase transition and this network evolves, emitting axions.  At the QCD phase transition each string becomes attached to $N$ domain walls. Throughout our discussion of Post-Inflation axion cosmology we assume a domain wall number of unity, $N=1$, so that there are no stable domain walls \cite{Sikivie:1982qv}.   It is an interesting question whether $N \neq 1$ universes are anthropically disfavored.  With $N=1$, at the QCD phase transition strings become the boundaries of domain walls, which collapse and disappear, again radiating axions.  Numerical simulations have studied the amount of axion cold DM arising from string evolution  \cite{Davis:1986xc, Hagmann:2000ja, Hiramatsu:2010yu} and  $N=1$ domain wall collapse \cite{Hiramatsu:2012gg}, reducing the value of $f$ needed to account for DM and introducing further uncertainties in $f$.  For illustration we adopt the range
\be 
f^{\text{Post}}  \, = \, (10^{10} - 10^{11}) \,\text{GeV}.
\label{eq:fPost}
\ee

The variables distinguishing between the two previous regimes, whether to average or not over the values of $\theta$, are the dynamics of inflation and reheating. In order for $a_i$ to be homogeneous within a Hubble patch the PQ phase transition has to have occurred already while the universe is inflating and the PQ symmetry must not be restored during reheating. These two facts require both $T_I\equiv H_I/2\pi$, the Gibbons-Hawking temperature of inflating de Sitter space, and $T_{\textrm{max}}$, the maximal temperature reached by the universe during reheating, to be smaller than $f$. If
\be
f> \max (T_I,T_{\max})
\ee
inflation will insure that $a_i$ is constant within our Hubble patch. This possibility allows the so called ``anthropic axion window''\footnote{The anthropic axion window \cite{Linde:1991km} does not require a multiverse, but follows from conventional inflation leading to large regions having different $\theta$ when the PQ phase transition occurs before inflation.  At large $f$ an anthropic requirement that $\xi_D$ not be too large selects for small $\theta$.  This anthropic requirement is much less well understood than the requirement of virialization used in this paper. } with larger values of $f$ but a small angle $\theta$.\footnote{An additional source of small scale fluctuation of the axion field are the quantum fluctuation which are imprinted on it during inflation. They determine a minimal value of the effective $\theta$ angle \cite{Turner:1990uz, Fox:2004kb}
\be
\theta_{\min}=\frac{H_I}{2\pi f}\approx 1.6\times 10^{-4}\left(\frac{H_I}{10^{12}\,{\textrm{GeV}}}\right)\left(\frac{10^{15}\,{\textrm{GeV}}}{f}\right)
\ee
and in turn a minimal contribution to $\rho_a$ in the Pre-Inflation case.} 

\begin{figure}
\begin{center} \includegraphics[width=0.65 \textwidth]{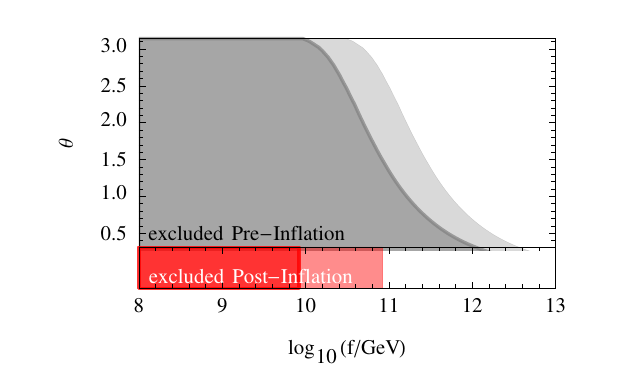}\end{center}
\caption{The gray and red shaded regions are excluded by the virialization requirement (\ref{eq:xic}) for Pre- and Post-Inflation axion cosmologies respectively, with light shading corresponding to the theoretical uncertainties of (\ref{eq:fPre}) and (\ref{eq:fPost}). \label{fig:f-theta}    }
\end{figure}

In Figure \ref{fig:f-theta} the shaded region in the $(f,\theta)$ plane is excluded by the virialization boundary (\ref{eq:xic}). In gray we plot the exclusion for the Pre-Inflation case in which the initial misalignment angle is constant in our Hubble patch. The thickness of the boundary of the excluded region shows the uncertainty from non-perturbative QCD corresponding to (\ref{eq:fPre}). The red shading corresponds to the exclusion for the Post-Inflation regime in which the misalignment angle is averaged over our Hubble patch. The theoretical uncertainty in this case (shown by the thick boundary of the excluded region) also includes a contribution from the axion density from radiation from axion strings and domain walls, and corresponds to the range of (\ref{eq:fPost}).

\begin{figure}[h!]
\begin{center}
\subfloat[]{\includegraphics[height=2.61in]{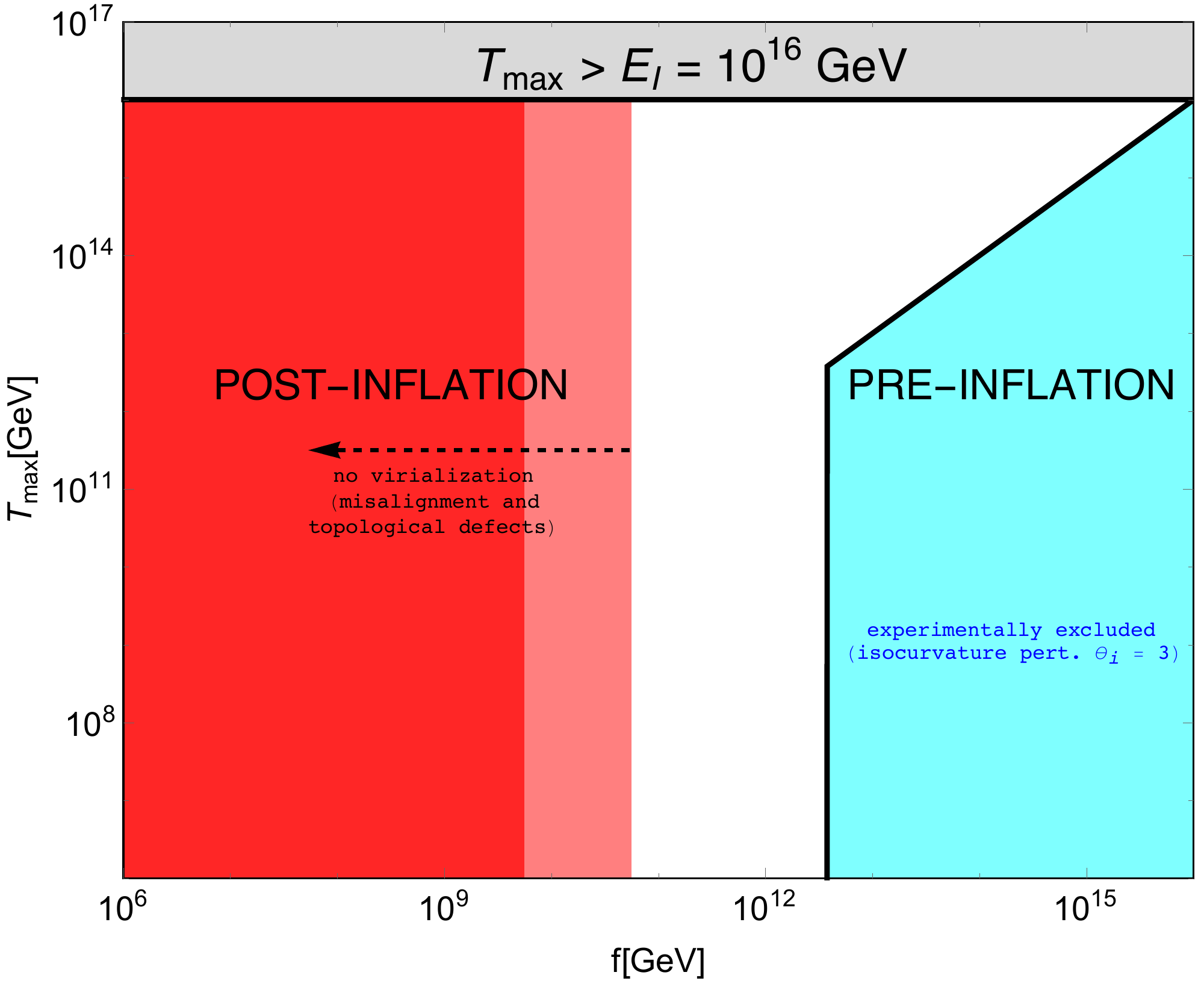}} $\quad$
\subfloat[]{ \includegraphics[height=2.61in]{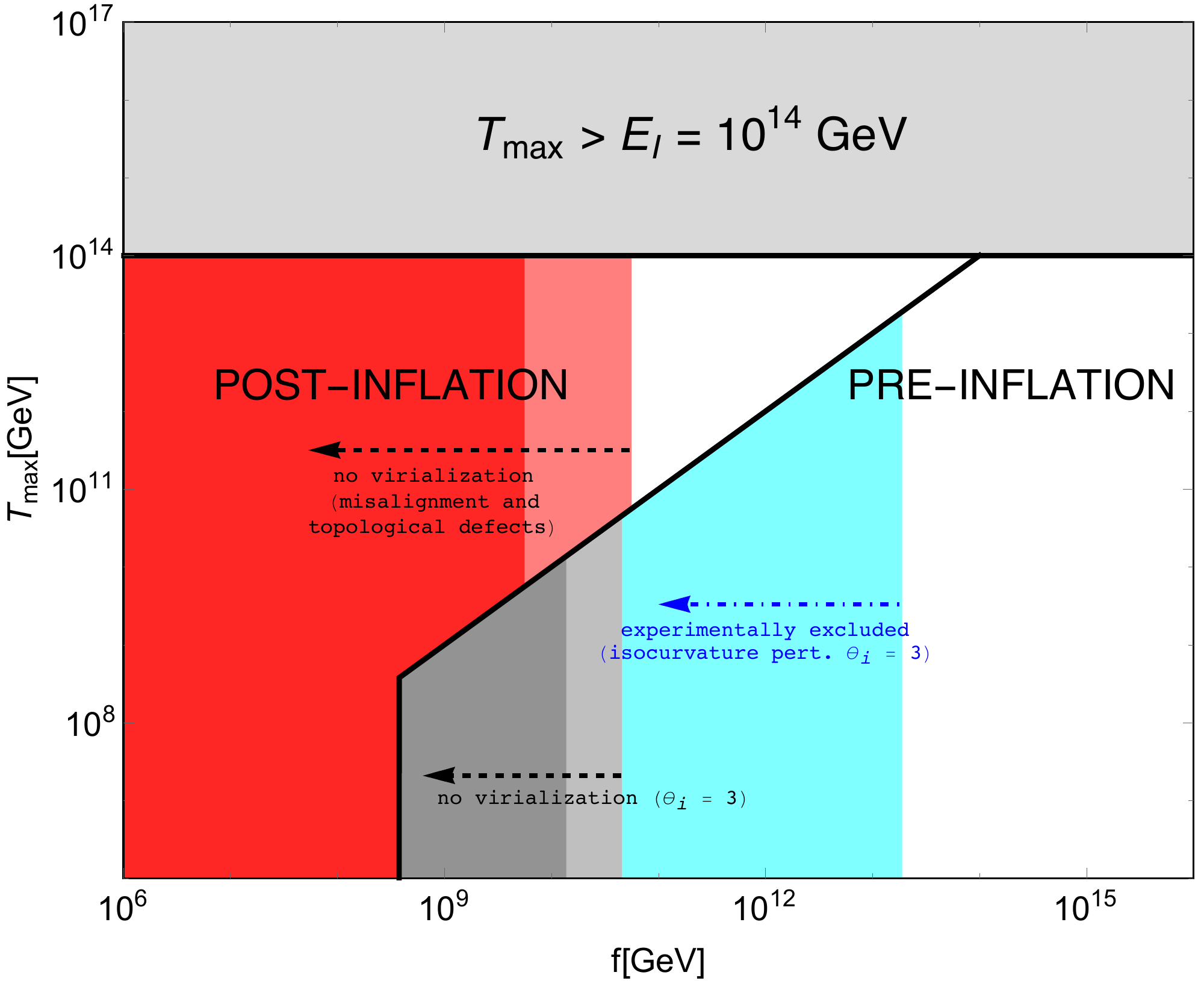}}  \\ 
\subfloat[]{ \includegraphics[height=2.61in]{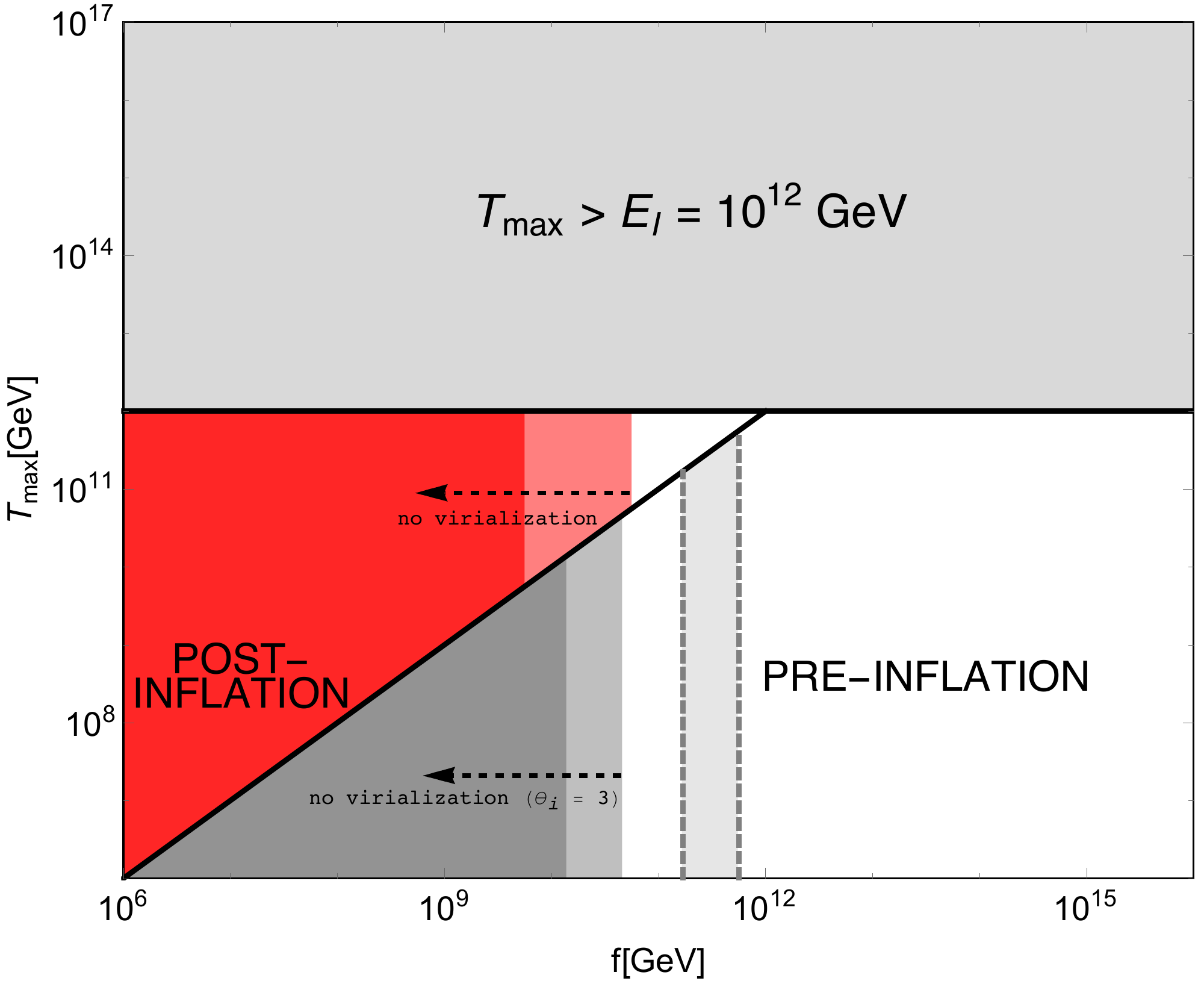}} $\quad$
\subfloat[]{ \includegraphics[height=2.61in]{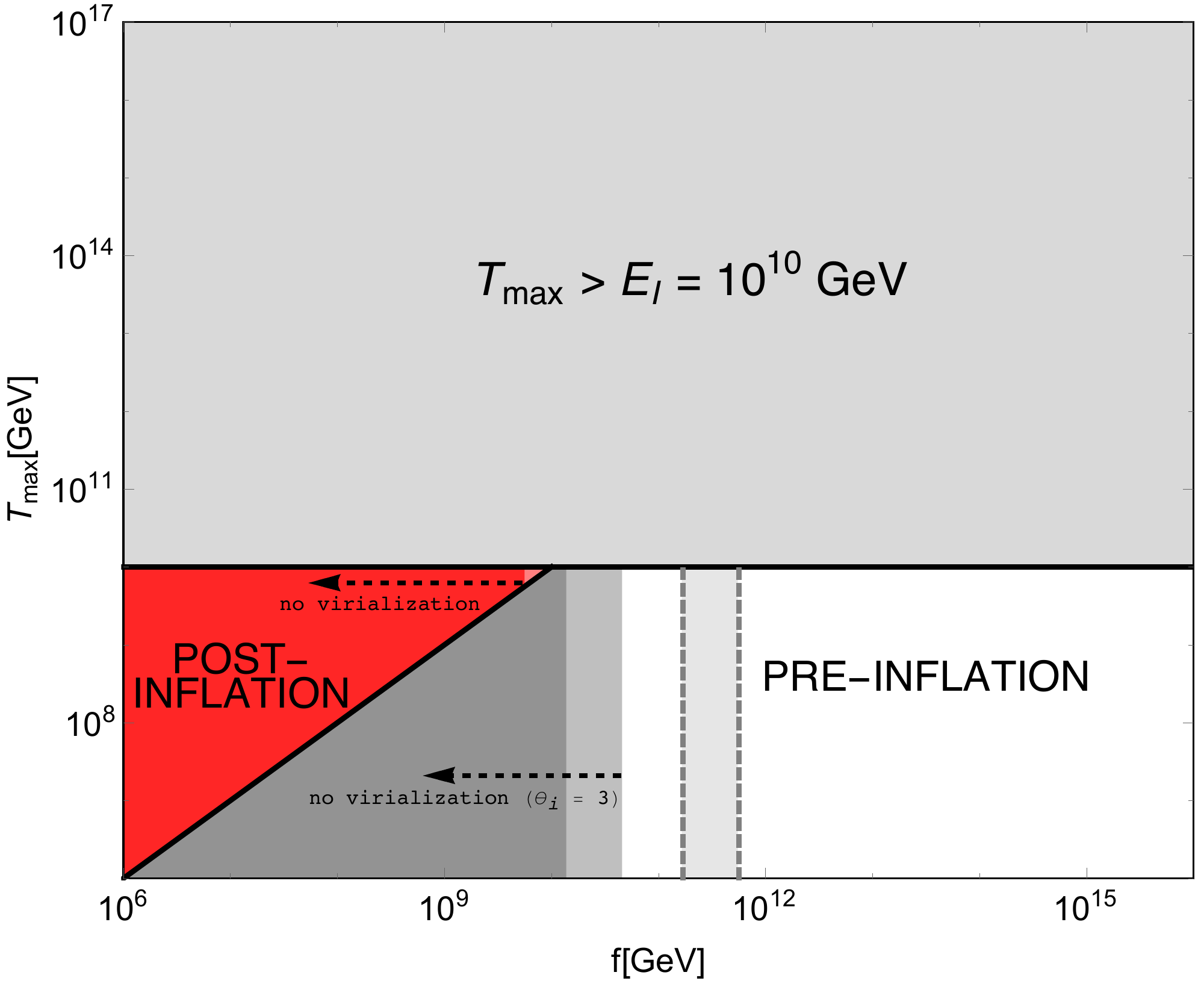}}  
\end{center}
\caption{The parameter space for axion DM.   Each panel shows the $(f, T_{\textrm{max}})$ plane for a fixed value of $E_I$.  The upper gray shaded regions are unphysical.  The solid black line separates regions with DM arising in the Post- and Pre-Inflation scenarios.  In the Post-Inflation regions the DM abundance is independent of $\theta$, while in the Pre-Inflation region it is proportional to $\theta^2 F(\theta)$.  The dark red (dark gray) shading shows  regions of Post- (Pre-) Inflation cosmology excluded by the virialization constraint, with the lighter shading corresponding to the uncertainty of  eq. \ref{eq:fPost} (\ref{eq:fPre}).  Pre-Inflation shading is shown for $\theta=3$, and in the lower two panels the light gray region with dashed boundaries shows the uncertainty band  for $\theta=1$. Blue shaded regions do not describe our universe as the isocurvature density perturbations are too large.  \label{fig:axionparam} }
\end{figure}

Figure \ref{fig:f-theta} does not yet capture the full parameter space of axion DM as it misses the effect of $T_I$ and $T_{\textrm{max}}$, the two parameters distinguishing between the Pre- and Post-Inflation scenarios. Their role is shown in Figure \ref{fig:axionparam}. In each of the four panels a different value for the energy scale of inflation, $E_I$, is chosen. Its relation to the inflationary Hubble parameter $H_I$ and to the Gibbons-Hawking temperature is through the Friedmann equation
\be
H_I=\sqrt{\frac{8\pi}{3}}\frac{E_I^2}{M_{Pl}}.
\ee
In each panel the physical region is the one in which the temperature at the end of inflation, $T_{\max}$, is smaller than $E_I$. The Pre- and Post-Inflation scenarios are separated by the boundary on which $f=\max (T_I,T_{\max})$.   The vertical part of the boundary has $f=T_I$, and moves to lower $f$ as $E_I$ is reduced from one panel to the next.  The remaining part of the boundary is the straight line of unit slope, $f=T_{\textrm{max}}$.  The red (gray) shaded regions are excluded by the virialization boundary (\ref{eq:xic}) in the Post- (Pre-) Inflation cosmology.  In each case the lighter shading corresponds to the theoretical uncertainty corresponding to (\ref{eq:fPost}, \ref{eq:fPre}).  For Pre-Inflation the position of the virialization boundary depends on $\theta$ and is shown for $\theta = 3$ in all panels and for $\theta = 1$ by the lightest gray shading in the lower two panels.

A further constraint in the Pre-Inflation scenario arises from the generation of axion isocurvature perturbations during inflation \cite{Fox:2004kb}; the regions shaded blue in Figure \ref{fig:axionparam} are observationally excluded for $\theta=3$. These perturbations grow rapidly with $E_I$, and at large $E_I$ there may also be anthropic constraints.


To explain $\xi_D$ from the virialization boundary we take an effective distribution that favors low $f$.  Figure \ref{fig:axionparam} then shows that the Post- (Pre-) Inflation cosmology results for larger (smaller) values of $E_I$ and $T_{\textrm{max}}$.  For $E_I < 10^{11}$ GeV, only the Pre-Inflation case is possible, while for $E_I > 10^{14}$ GeV the description of our universe by Pre-Inflation cosmology becomes less probable.  

\subsection{Scanning over $f$ and $\theta$}

In the previous sub-section we have shown the regions of axion parameter space excluded by the anthropic requirement that DM density perturbations go non-linear, allowing virialization and halo formation. 
As discussed in the introduction, in this paper we explore the possibility that the observed DM energy density is determined by this virialization boundary, taking the cosmological constant to be anthropically constrained by observer dilution, as illustrated in Figure \ref{fig:Lambda-xi}.

We now investigate allowing $f$ to scan over the multiverse according to a given prior probability distribution
\be
dP \propto p(f) d\ln f.
\ee
We stress that this is rarely studied, and is not the case in the conventional ``anthropic window" where only $\theta$ scans.\footnote{Ref.~\cite{hertzberg} also allows the value of $f$ to scan to investigate possible relations between axion physics and instabilities in the SM Higgs potential.}  In the Pre-Inflation case the initial value of the misalignment angle is an additional parameter, with a flat probability distribution between $-\pi$ and $\pi$.   From Figure \ref{fig:Lambda-xi} we know that $\xi_D$ is close to the virialization boundary, and from Figure \ref{fig:f-theta} the form of the virialization boundary requires that $p(f)$ increases towards low $f$ in the vicinity of the boundary.   We assume, for simplicity, that the distribution can be approximated as a  power law in the vicinity of the boundary, $p(f)=f^n$.  We study which values of $n$ lead to our proximity to the boundary and, in the Pre-Inflation case where $\theta$ is also scanning, we obtain the corresponding probability distributions for $f$ that reproduce the observed DM abundance.  

In order to obtain the posterior probability distribution for $f$ we have to take into account measure and anthropic selection effects which modify the above simple power law behavior, and recall that our overall scheme includes the scanning of $\Lambda$.  Following the discussion in section \ref{sec:xiddist} that led to (\ref{eq:Pinobs}) and (\ref{eq:xic}), after marginalizing over $\Lambda$ the effective probability distribution for $f$ and $\theta$ becomes
\be
\label{measuref}
dP \, \propto \, \theta(\xi_D-\xi_c) \, \frac{1}{1+\xi_D/\xi_{B0}} f^n \, d\ln f \; (d \theta)
\ee
where the virialization boundary is approximated by a $\theta$ function at $\xi_c = 0.5 \, \xi_{D0}$. 
The integral $(d\theta)$ is present only for the Pre-Inflation case.

\begin{figure}
\begin{center}
\includegraphics[height=3.5in]{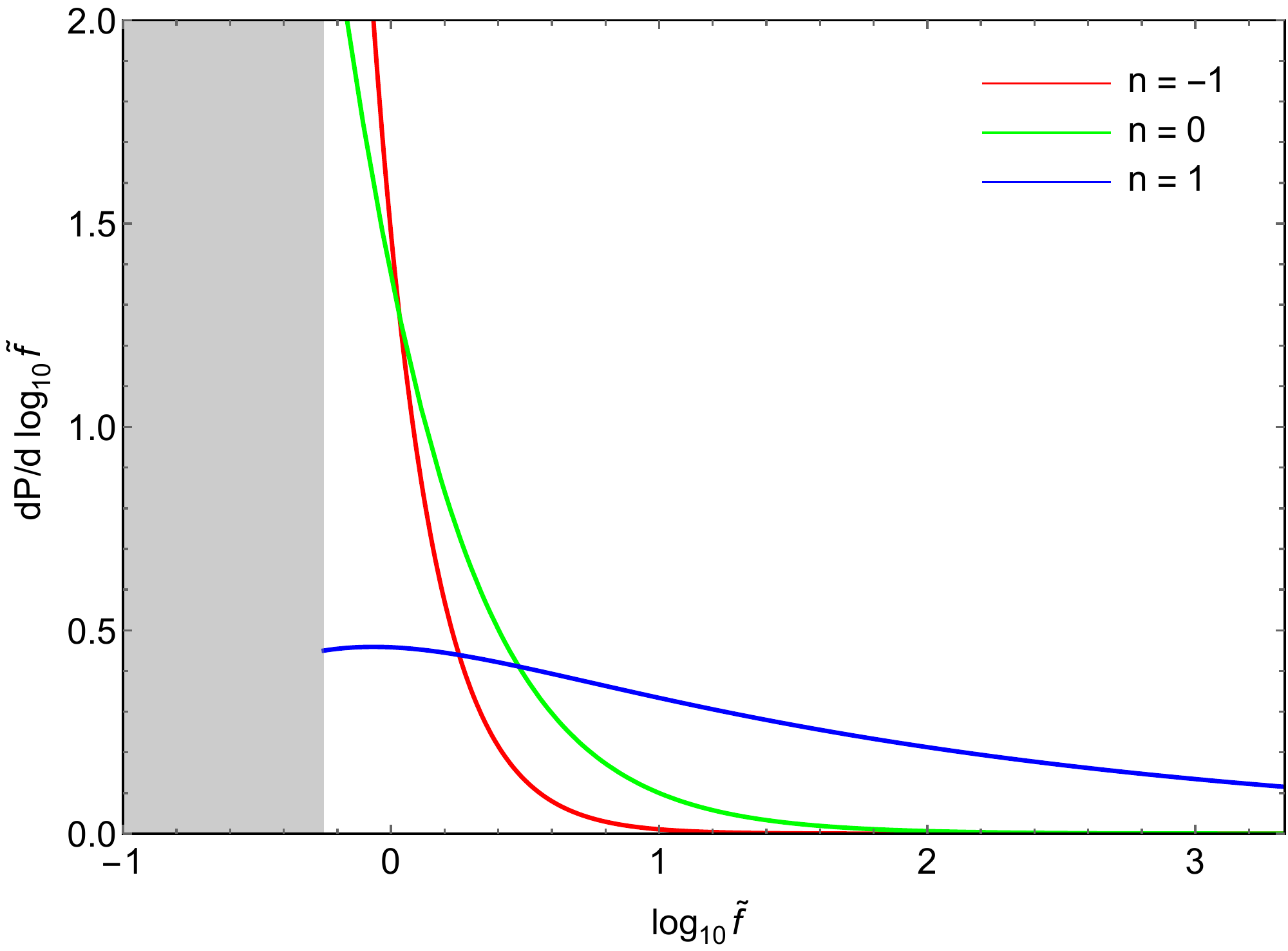}
\end{center}
\caption{Post-Inflation axion DM: probability distributions for $\tilde{f} = f/f_0$ for three different values of $n$. }
\label{fig:ProbDistr1}
\end{figure}

For the Post-Inflation cosmology, we parametrize
\be
\xi_D/\xi_{B0}  = 6.0 \, \tilde{f}^{\,1.2} \qquad{\textrm{with}}\qquad \tilde{f} = f/f_0\,,
\label{eq:axionrelic2}
\ee
where $f_0$ is the value for which the observed DM abundance is reproduced and has a theoretical uncertainty from the contribution of axion topological defects. 
Using the variable $\tilde{f}$ the probability distribution reads
\be
dP \propto \, \theta(\tilde{f} - \tilde{f}_c) \,   \frac{1}{1+ 6.0\tilde{f}^{\,1.2}} \; \tilde{f}^{\,n} \, d \ln \tilde{f}
\label{eq:probPOST}
\ee
where $\tilde{f}_c \simeq 0.56$ is the value of $\tilde f$ corresponding to $\xi_D=\xi_c$. In order to get a normalizable distribution for $n>1.2$ we cut off the range of $\tilde f$ at $\tilde{f}_{\rm max} = 2.15 \times 10^3$, correspondent to $\xi_D = \xi_{\rm max} = 10^4 \, \xi_{D0}$. This can be interpreted as an additional anthropic boundary at large $\xi_D$ related for instance to close stellar encounters \cite{Linde:1985gh, Tegmark:2005dy}. The probability distributions for three different values of $n$ are shown in Figure \ref{fig:ProbDistr1}.

Turning to the Pre-Inflation scenario, the probability distribution is doubly differential, with also the initial misalignment angle scanning uniformly between $-\pi$ and $\pi$.  Hence $d\theta$ is included in (\ref{measuref}) and the axion abundance in this case is given by (\ref{rhoaxion}). The 2-dimensional distribution in the $(\theta, f)$ plane is not particularly illuminating. An important question to answer is about the support of the distribution in this case. We trade the variable $f$ for $\xi_D$, and as already discussed the DM abundance scans over the interval
\be
\xi_c < \xi_D < \xi_{\rm max} = 10^4 \ .
\label{eq:xidom}
\ee
We require that we never scan over values of $f$ greater than the Planck mass, which translates into the condition
\be
\theta_{\rm min} = 10^{-2} < \theta < \pi \ , 
\label{ed:thetadomain}
\ee
with $\theta_{\rm min}$ determined by $\xi_D(\theta_{\rm min}, f = M_{\rm Pl}) = \xi_{\rm max}$. 

\begin{figure}
\begin{center}
\includegraphics[height=3.3in]{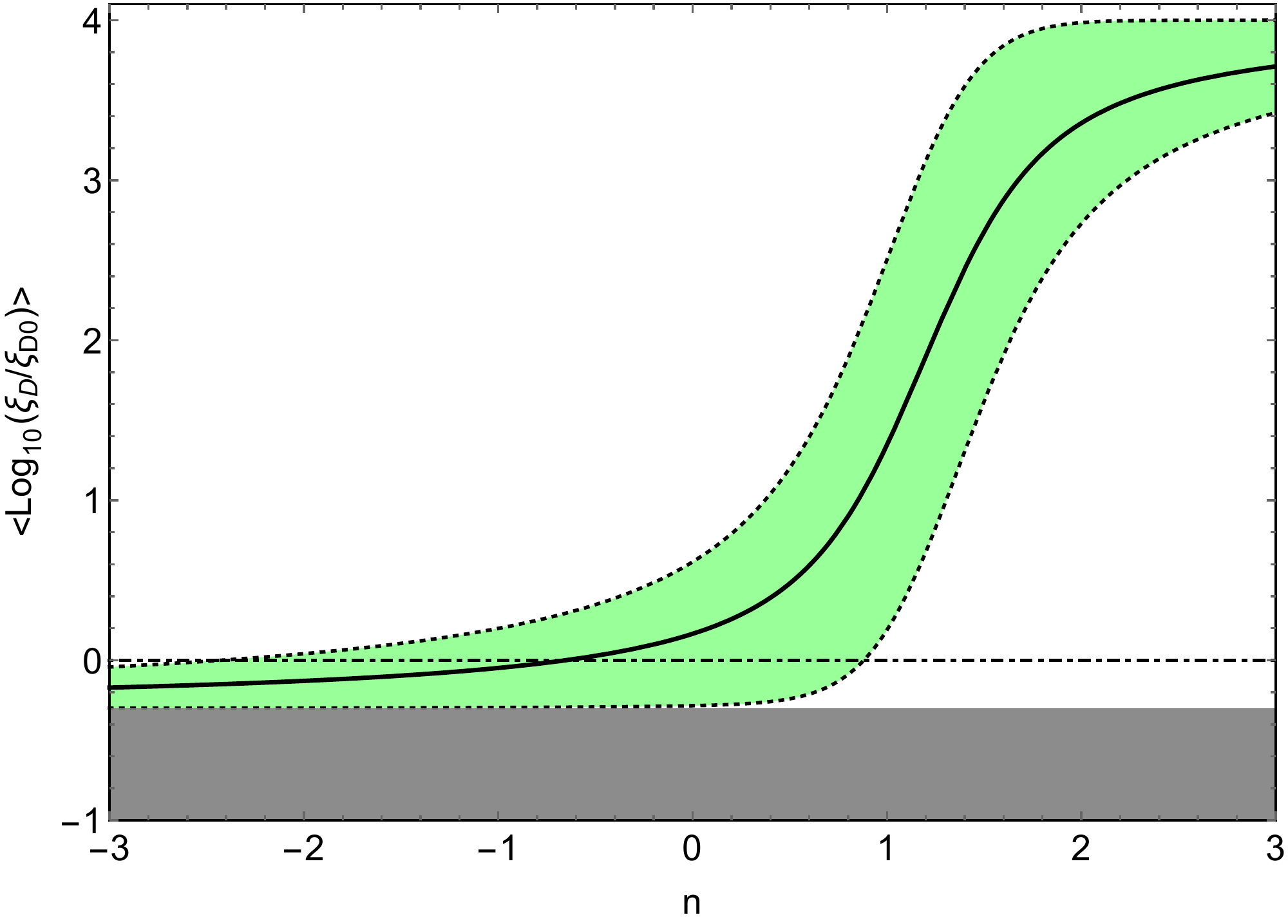}
\end{center}
\caption{Average (solid black line) and $1\sigma$ range (green band) of $\xi_D$ as a function of $n$. The result is valid for both Post-Inflation and Pre-Inflation scenario. The DM abundance observed is our universe is reproduced at $1\sigma$ for $- 2.42 \leq n \leq 0.88$.}
\label{fig:ProbDistr2}
\end{figure}

We use the probability distributions to compute the average value of $\xi_D$ as well as the $1 \sigma$ confidence interval for both cosmologies, and we discuss the range of $n$ that makes the observed abundance of DM in our universe typical. For Post-Inflation cosmology we change variable in the distribution (\ref{eq:probPOST}) by using \Eq{eq:axionrelic2}, and we derive a probability distribution for $\xi_D$. For Pre-Inflation cosmology we start from the double differential distribution in \Eq{measuref}, trade $f$ for $\xi_D$ using \Eq{rhoaxion} and then compute the average of $\xi_D$ in the domain described by \Eqs{eq:xidom}{ed:thetadomain}. The additional integration over $\theta$ in the Pre-Inflation case can be performed independently, thus in the two cases we have the same results, which are shown in Figure \ref{fig:ProbDistr2}. We find that our universe is `$1\sigma$-typical' for $- 2.42 \leq n \leq 0.88$. It is worth emphasizing that this range is not sensitive to the detailed choice of $\xi_{\rm max}$, and it stays unaffected for $\xi_{\rm max} = (10^3 \div 10^6) \, \xi_{D0}$.

\begin{figure}
\begin{center}
\includegraphics[height=3.4in]{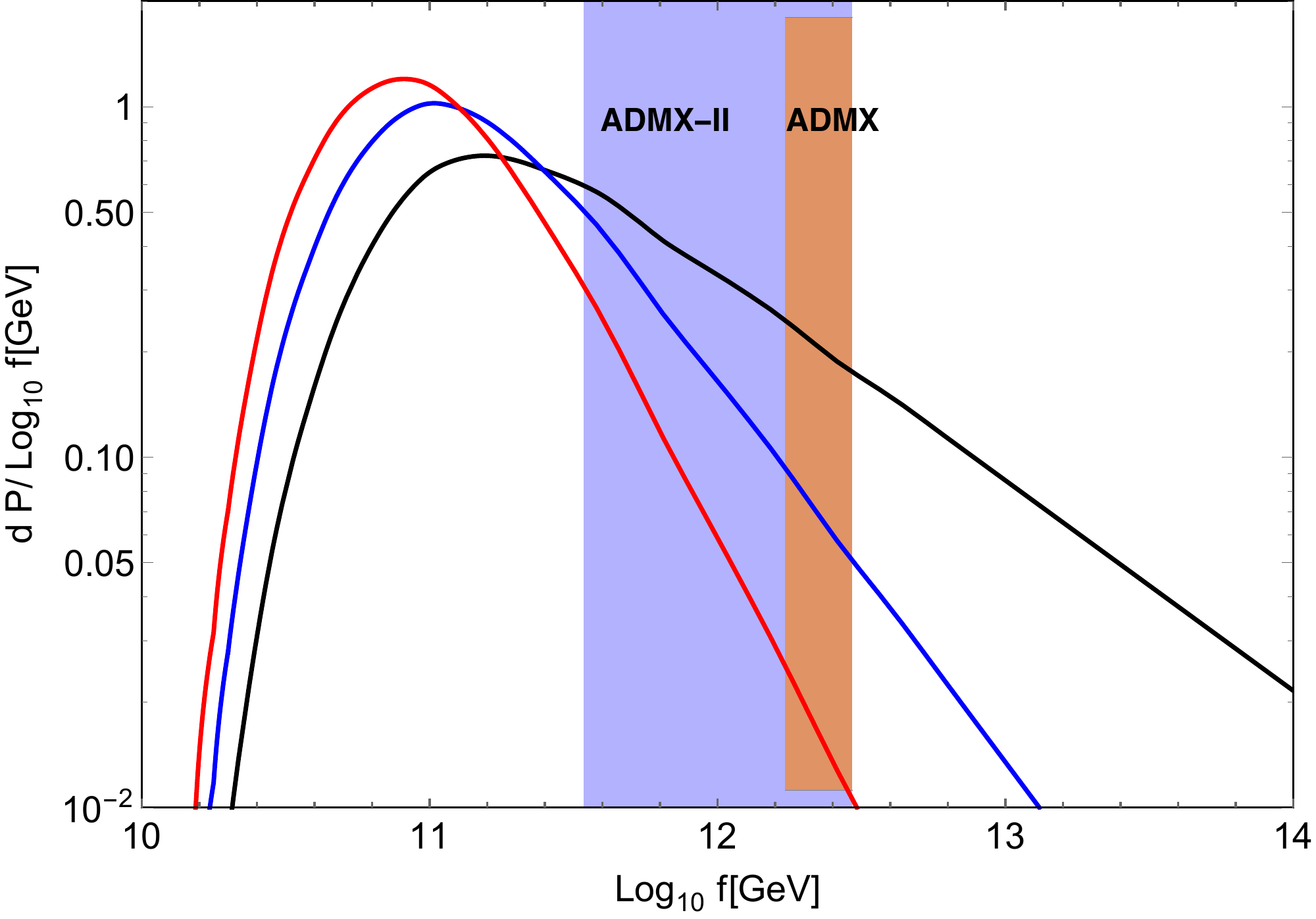}
\end{center}
\caption{Probability distributions for $f$ in the Pre-Inflation scenario imposing $\xi_D=\xi_{D0}$.}
\label{fig:ProbabilityPREfull}
\end{figure}

\begin{figure}
\begin{center}
\includegraphics[height=3.4in]{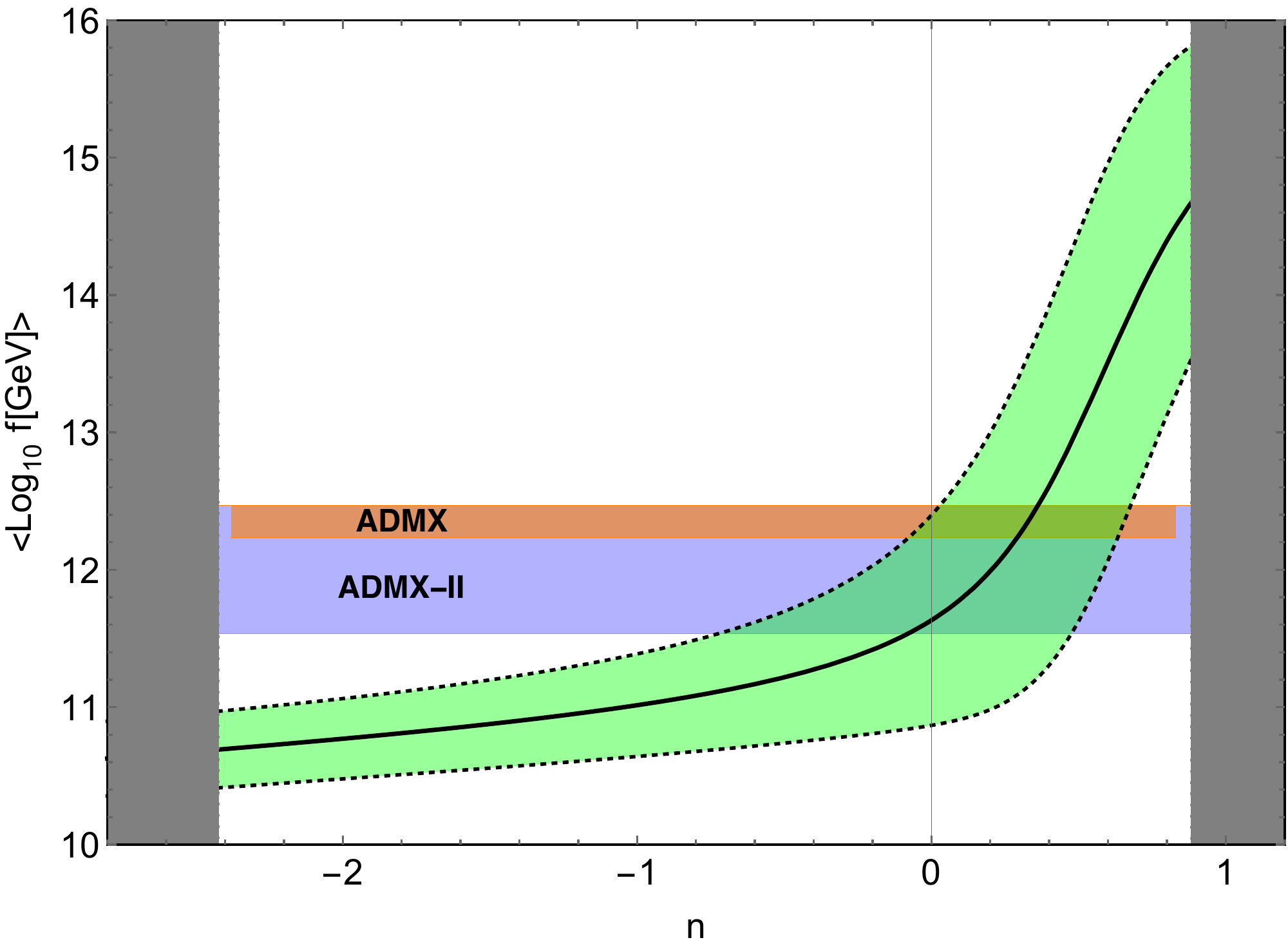}
\end{center}
\caption{Average (solid black line) and $1\sigma$ range (green band) of $f$ as a function of $n$ in the Pre-Inflation scenario.}
\label{fig:AverageF}
\end{figure}

In the Pre-Inflation case, for each value of $n$ that successfully gives our proximity to the virialization boundary, we can obtain the probability distribution for measuring $f$ in our universe given the observed value of $\xi_{D0}$.  We thus proceed to integrate out $\theta$ from (\ref{measuref}) imposing that the total axion density is the one observed today. The probability distribution for $f$ thus reads
\be\label{pdfADMX}
dP \propto f^n \,d\ln f\int_{-\pi}^\pi\delta(\xi_D-\xi_{D0}) \, \frac{1}{1+\xi_D/\xi_{B0}} \, d\theta.
\ee
Using (\ref{rhoaxion}) for $\xi_D$ and including anharmonic effects, numerical integration over $\theta$ yields the
distributions of Figure \ref{fig:ProbabilityPREfull}, shown for three values of $n$.  More generally, Figure \ref{fig:AverageF} shows the average and $1\sigma$ ranges of $f$ for all values of $n$ that give our proximity to the virialization boundary at $1\sigma$.  We caution the reader that this green shaded $1\sigma$ region contains uncertainties due to our approximation of the virialization boundary as a $\theta$ function at $\xi_c$.  The messages from Figures \ref{fig:ProbabilityPREfull} and \ref{fig:AverageF} are clear.  For a wide range of $n$ that give our universe close to the virialization boundary, $-3 < n < 0$,  the expected values of $f$ are small, centered on $(0.3 - 3) \times 10^{11}$ GeV.  Only the smaller range of $n$ above zero leads to the expectation of larger values of $f$.   The case $n=0$ is particularly interesting, occurring if the PQ scale is generated dynamically via a dimensional transmutation, and yields $f \approx (10^{11} - 10^{12})$ GeV.

One interesting question to ask is how well an experiment like ADMX would perform in discovering the axion assuming it is the DM. The answer to this question depends on whether we live in the Pre- or Post-Inflation scenario and on the value of $n$. According to the experimental collaboration, the ADMX and ADMX-II experiments are going to be sensitive to the following ranges of $f$ 
\be
\begin{split}
{\textrm{ADMX:}} & \, \qquad 1.7 \times 10^{12} \, \GeV \lesssim f \lesssim 3 \times 10^{12} \, {\rm GeV} \ , \\ 
{\textrm{ADMX-II:}} & \, \qquad 3.4 \times 10^{11} \, \GeV \lesssim f \lesssim 3 \times 10^{12} \, {\rm GeV} \ , \\ 
\end{split}
\ee
which are shown as shaded bands in Figures \ref{fig:ProbabilityPREfull} and \ref{fig:AverageF}.

Unfortunately neither of these two phases of the experiment is expected to cover a Post-Inflation axion, where $1.6\times 10^{10}$\, GeV$\lesssim f\lesssim 1.6\times 10^{11}$\,GeV depending on whether the contribution from the decay of topological defects is taken into account or not.\footnote{However, ADMX-HF is a second platform specifically designed to reach lower $f$ \cite{Shokair:2014rna}. Also see the recent proposals in \cite{Arvanitaki:2014dfa}, \cite{Sikivie:2014lha}, \cite{Stadnik:2013raa} and \cite{Roberts:2014dda} .}

\begin{figure}
\begin{center}
\subfloat[]{\label{fig:ProbADMXI} \includegraphics[height=2.6in]{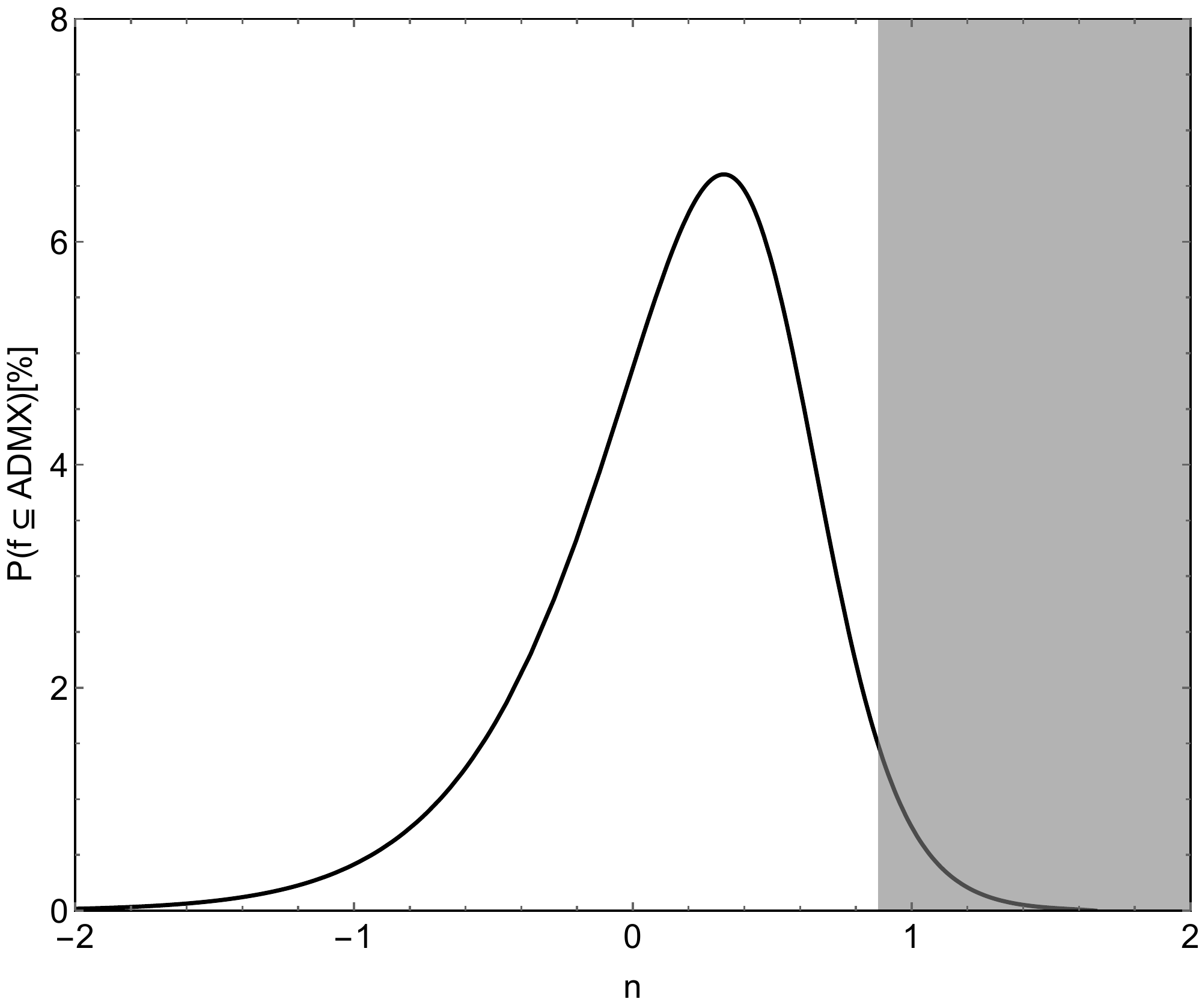}} $\qquad$
\subfloat[]{\label{fig:ProbADMXII} \includegraphics[height=2.6in]{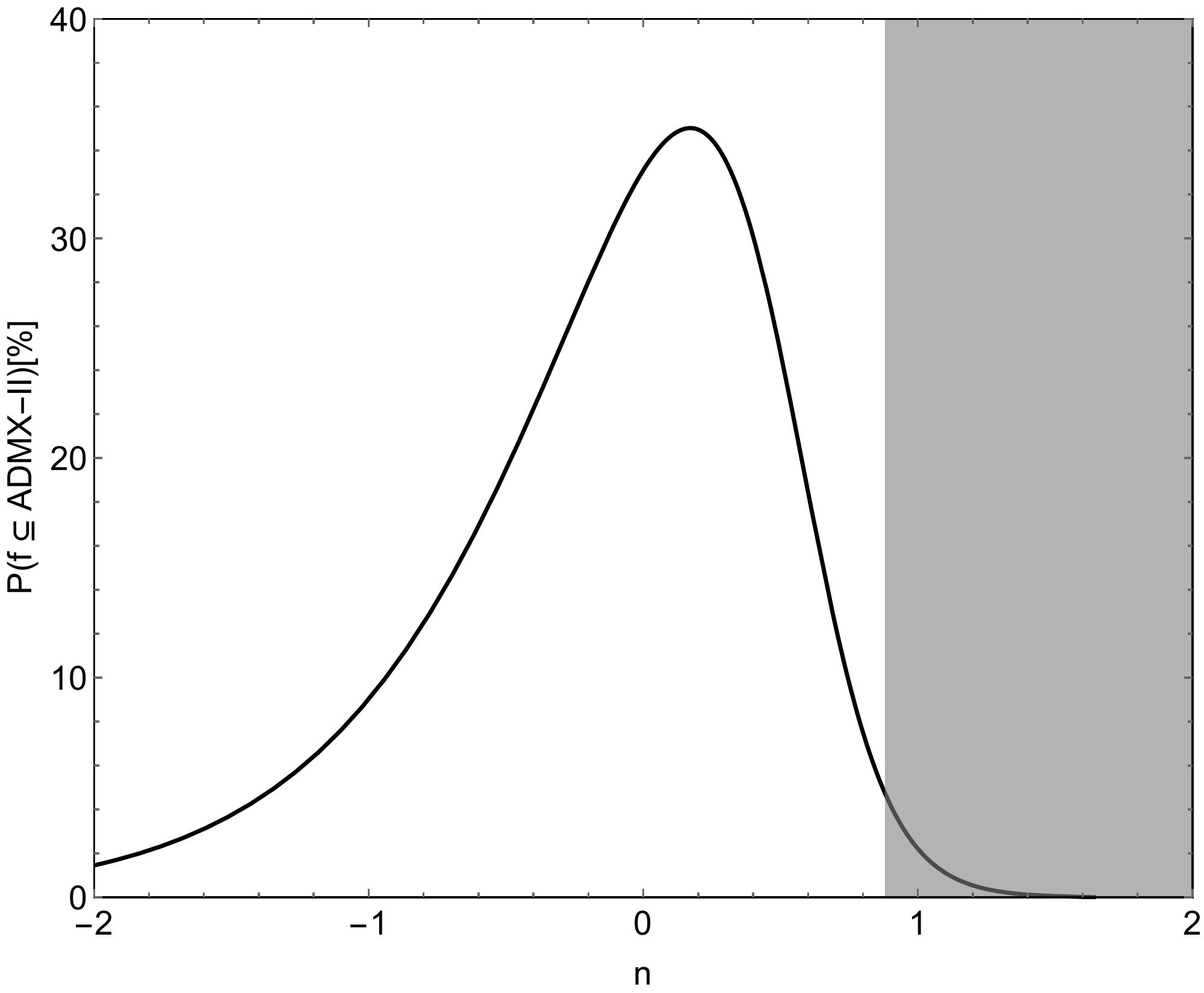}} 
\end{center}
\caption{Probability that $f$ is in the region probed by ADMX (left) and ADMX-II (right) as a function of $n$. The gray region corresponds to the values of $n$ for which our universe is not `$1\sigma$-typical'.}
\label{fig:ProbADMX}
\end{figure}

The situation is different in the Pre-Inflation case, where a larger value of $f$ can be accommodated by a small initial misalignment angle $\theta$.  However, the distribution for $f$ is constrained by requiring our universe to be near the virialization boundary, and Figures \ref{fig:ProbabilityPREfull} and \ref{fig:AverageF} show that, for a wide range of allowed $n$, $f$ is typically below the reach of these experiments.   The probability for $f$ to be in the (ADMX, ADMX-II) reach is shown in Figure \ref{fig:ProbADMX}(a,b).   Thus for $n$ not too far from zero there is hope that these experiments will make a positive discovery, but this should not detract from the intense need to design experiments to probe the lower $f$ region.



\subsection{The Thermal Axion Window}
\label{subsec:thermal}

The observed DM abundance can be understood if the multiverse favors low values of $f$. If this is the case, we live close to the catastrophic boundary coming from requiring sufficient DM for density perturbations to go non-linear and halos to form by virialization.   However, the argument presented in this Section is not quite complete because, for low enough $f$, sufficient axions are produced from thermal scattering for the axion DM density to rise above the virialization bound.  This low $f$ region is observationally excluded, for example from limits on axion emission from supernovae and from white dwarfs; we argue now that it is also anthropically disfavored. 

The calculation of the density of thermal axions is performed in App.~\ref{app:thermal}. Skipping all the calculation details, we quote the final result for the thermal axion density, which arises from hadron collisions below the QCD phase transitions
\be
\Omega_a h^2 \simeq  \, 0.12 \, \frac{3.4 \times 10^5 \, \GeV}{f}  \ .
\ee
In order for this region not to be anthropically excluded from virialization, the axion density must above the catastrophic boundary $\xi_a > \xi_c$, so that $f$ must be below a critical value
\be
f < f^{th}_c \simeq 7 \times 10^5 \, {\rm GeV} \ .
\ee

\begin{figure}
\begin{center}
\includegraphics[height=3.3in]{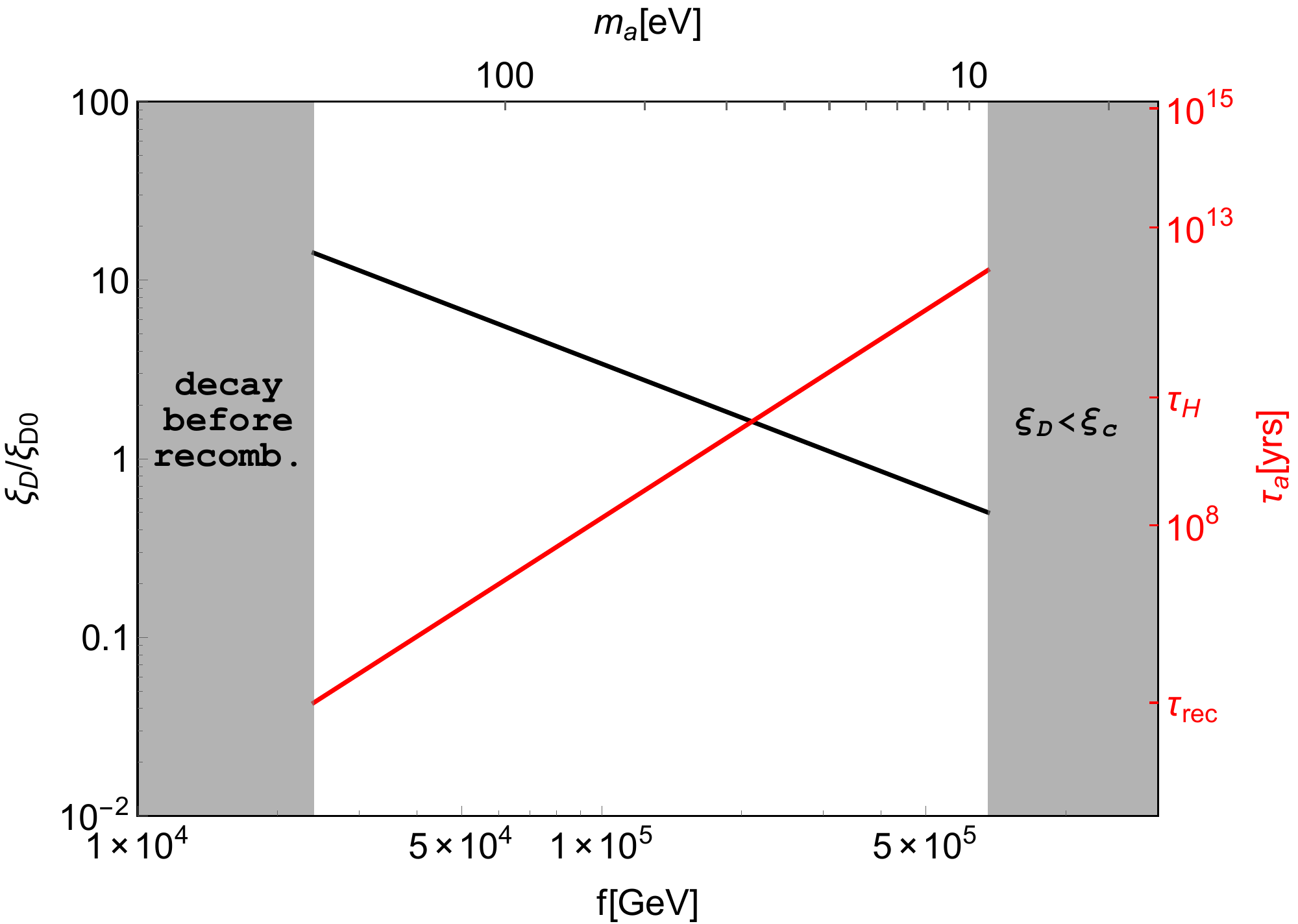}
\end{center}
\caption{Energy density for thermal axions (black line) and axion lifetime (red line) as a function of the PQ breaking scale (lower axis) and the axion mass (upper axis). In the shaded regions density perturbations do not go non-linear, preventing halo formation.}
\label{fig:ThermalAxions}
\end{figure}

The PQ breaking scale cannot be arbitrarily small, and the lower limit is set by a catastrophic boundary coming again from large scale structure. Small values of $f$ make the axion strongly coupled to SM fields, and very short lived through the decay process $a \rightarrow \gamma \gamma$. If there is no DM around at recombination, baryons would not be able to fall into DM potential wells, and therefore all the perturbations below the Silk scale would be damped. This translates into the anthropic condition $\tau_{a \rightarrow \gamma \gamma} \geq t_{\rm rec}$, with $t_{\rm rec}$ the time when recombination happens. The axion life-time can be computed from the effective lagrangian in (\ref{effaxion}), and it results in
\be
\tau_{a \rightarrow \gamma \gamma} \simeq 
\frac{1.3 \times 10^{13}}{c_\gamma^2} \left(\frac{f}{10^6 \, \GeV}\right)^5 \, {\rm yrs} \ .
\ee
The coupling $c_\gamma$ is model-dependent (see e.g.~\cite{Georgi:1986df}), but typically of order one, so we set $c_\gamma = 1$ in what follows. The recombination temperature is $T_{\rm rec} \simeq 0.3 \, {\rm eV}$, for all purposes independent of the DM abundance. We use the Friedmann equation to obtain $t_{\rm rec}$; for $f \leq f^{th}_c$ the axion is non-relativistic at recombination and the equivalence temperature satisfies $T_{\rm eq} \geq T_{\rm rec}$, therefore we consider a matter dominated universe. We find that this imposes $f > f^{th}_{\rm rec}$, where
\be
f^{th}_{\rm rec} = 2.4 \times 10^4 \, {\rm GeV} \ .
\ee

There is indeed a thermal window where axion DM would not be excluded by virialization: $f^{th}_{\rm rec} \leq f \leq f^{th}_c$. This low-$f$ region is shown in Figure \ref{fig:ThermalAxions}, where we plot both the axion relic density and the axion life time as a function of $f$ (and the axion mass, upper axis). The shaded gray areas are excluded by virialization, but the region between them is in principle viable. Interestingly, the value of $f$ such that the axion lifetime is of the order of the Hubble time falls into this range, $f_H \simeq 2.5 \times 10^5 \, \GeV$.

\begin{figure}
\begin{center}
\includegraphics[height=3.3in]{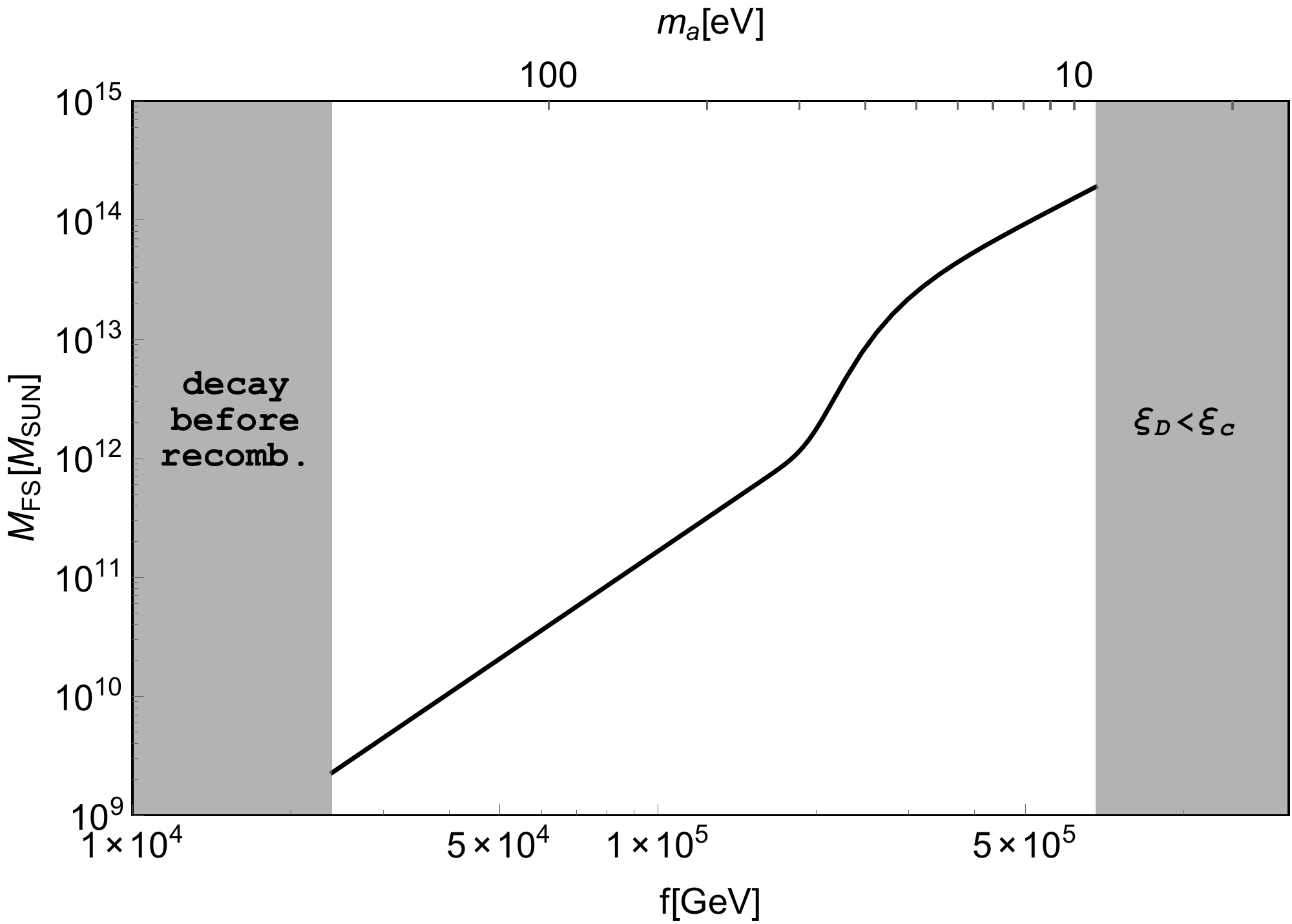}
\end{center}
\caption{Free streaming mass scale for thermal axions as a function of the PQ breaking scale (lower axis) and the axion mass (upper axis).}
\label{fig:AxionFreeStreaming}
\end{figure}

There are anthropic reasons why this region is unlikely. Thermal axions produced for such low values of $f$ are hot DM, and after they decouple they can free stream from overdense to underdense regions, damping primordial perturbations. The free streaming scale $\lambda_{\rm FS}$ can be estimated by the expression (see e.g.~\cite{Kolb:1990vq})
\be
\lambda_{\rm FS} \simeq 20\, {\rm Mpc} \frac{10 \, {\rm eV}}{m_a}  \ .
\ee
This length has an associated mass, corresponding to the mass scale of the perturbation which are free streamed away, resulting in
\be
M_{\rm FS} = \rho_{\rm FS} \, \frac{\pi}{6} \, \lambda_{\rm FS}^3  \ .
\ee
The mass density $\rho_{\rm FS}$ has to account for the fact that axions can be unstable on a cosmological scale, thus we use the expression
\be
 \rho_{\rm FS}(f) = \rho_B + \rho_a\, e^{- \frac{t_H}{\tau_{a \rightarrow \gamma \gamma}} } \ .
\ee
The final result is shown in Figure \ref{fig:AxionFreeStreaming} and implies that, at the very least, large scale structure formation is very different from our universe.  Throughout the thermal axion window perturbations on scales $M<10^9 M_\odot$ are destroyed by axion free streaming, so that population III stars fail to form.  In about half of the window even galaxies of $M= 10^{12} M_\odot$ fail to form.


\section*{Acknowledgments}

We thank Raphael Bousso and Sergei Dubovsky for useful comments and discussions. DP thanks Richard P. Gomez for discussions.
This work was supported in part by the Director, Office of Science, Office 
of High Energy and Nuclear Physics, of the US Department of Energy under 
Contract DE-AC02-05CH11231 and by the National Science Foundation under 
grants PHY-1002399 and PHY-1316783. L.H. thanks the hospitality of the Aspen Center for Physics (NSF Grant PHY-1066293), where part of this work was carried out.  F.D. is supported by the Miller Institute for Basic Research in Science.

\appendix

\section{Boltzmann Equation Analysis for Thermal Axions}
\label{app:thermal}

In this Appendix we give the details of how to compute the relic density of thermal axion. The evolution of the axion number density $n_a$ is governed by the Boltzmann equation
\be
\frac{d n_a}{dt} + 3 H n_a = - \Gamma_a \, \left[n_a - n^{\rm eq}_a \right] \ .
\label{eq:BoltzAxion}
\ee
Here, $H$ is the Hubble parameter, $\Gamma_a$ the rate for reactions producing axions and $n_a^{\rm eq}$ the axion equilibrium density. The $3 H n_a$ factor accounts for the dilution due to the expansion, the right hand side accounts for interactions changing the axion number. The rate is dominated by $2 \leftrightarrow 2$ reactions of the form $i j \leftrightarrow k a$ changing the axion number
\be
\Gamma_a = \sum_{i,j,k} \Gamma_a(i j \leftrightarrow k a) \ .
\ee 
Each individual reaction gives the contribution
\be
\Gamma_a(i j \leftrightarrow k a) = \frac{1}{n_a^{\rm eq} } \int d\Pi_i d\Pi_j \,  f^{\rm eq}_i f^{\rm eq}_j  \, \left|\mathcal{M}_{ij \rightarrow ka}\right|^2 \, (2\pi)^4 \delta^4\left(p_i+p_j-p_k-p_a\right) \, d\Pi_k d\Pi_a \ ,
\label{eq:axionrate}
\ee
where we introduce the relativistic invariant phase space
\be
d \Pi_i = g_i \frac{d^3 p_i}{2 E_i (2\pi)^3} \ .
\ee

We want to study whether axion thermalization is achieved, therefore it is practical to use the variable $\eta = n_a / n_a^{\rm eq}$ to track the axion density. It is also convenient to trade the time $t$ with the temperature $T$. The Boltzmann equation in these new variables takes the form
\be
T \, \frac{d \eta}{d T}  =  \left(1 + \frac{1}{3} \frac{d \log g_{*s}}{d \log T}\right) \frac{\Gamma_a}{H} \left(\eta - 1\right) +
\eta \frac{d \log g_{*s}}{d \log T} \ ,
\label{eq:BoltzGlobal2}
\ee
with $g_{*s}$ the effective number of relativistic degrees of freedom contributing to the entropy density. The boundary condition for the Boltzmann equation is $\eta(f) = 0$, since there are no axions above the PQ breaking scale.

\begin{figure}
\begin{center}
\subfloat[]{\label{fig:Tdec1} \includegraphics[height=2.5in]{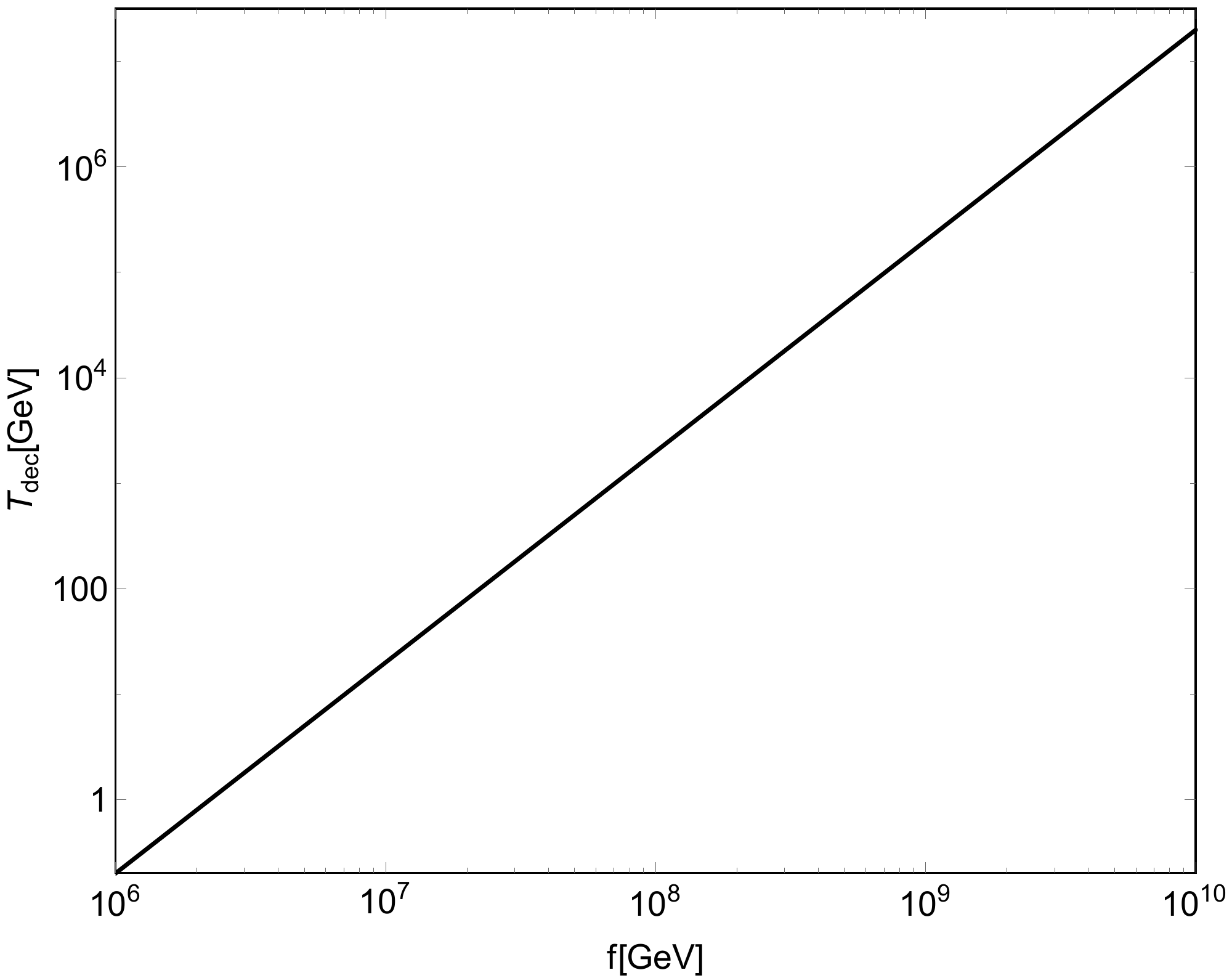}} $\qquad$
\subfloat[]{\label{fig:Tdec2} \includegraphics[height=2.5in]{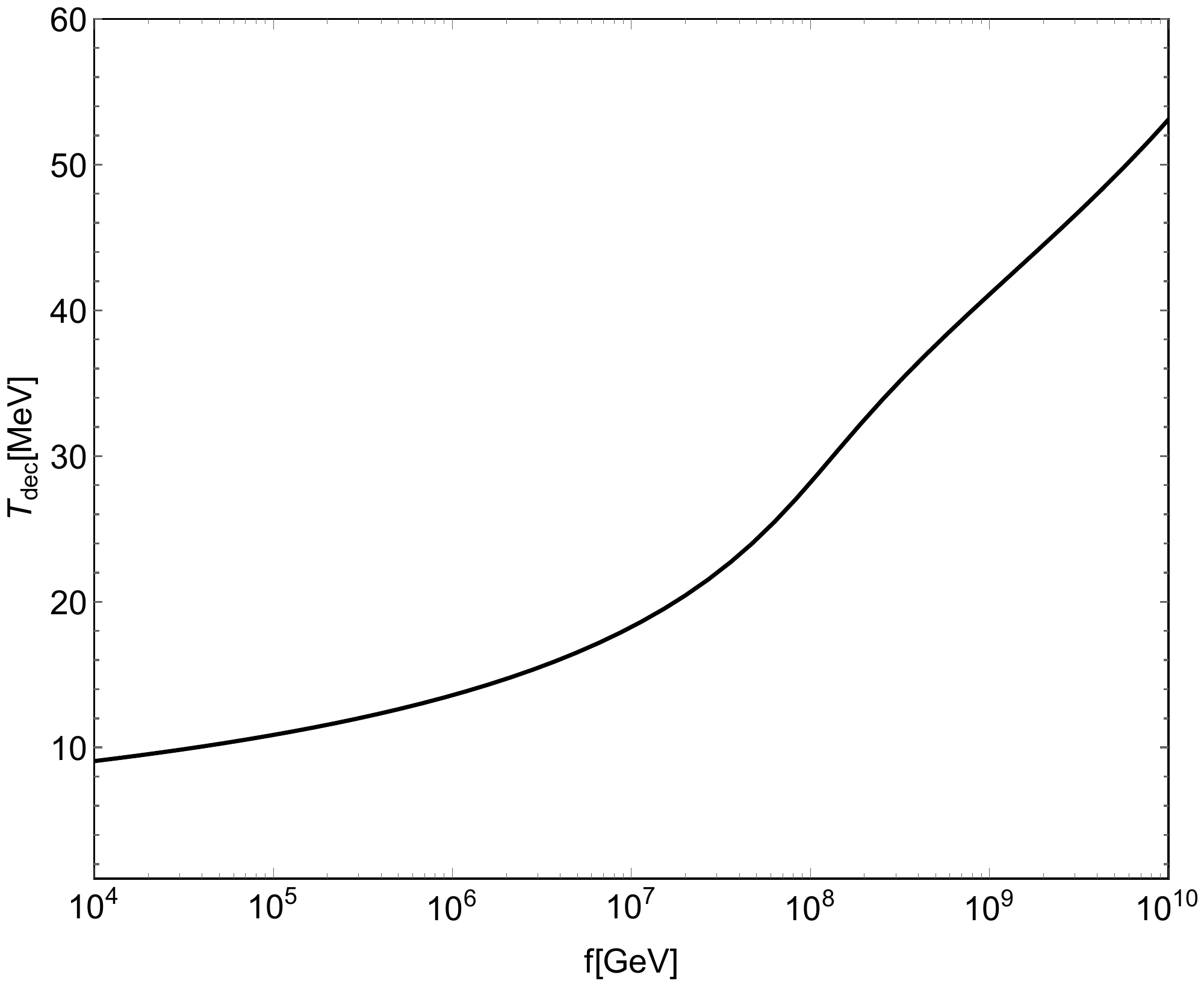}}  
\end{center}
\caption{Decoupling temperature as a function of the PQ breaking scale $f$: (a) above the QCD phase transition; (b) below the QCD phase transition.}
\label{fig:Tdec}
\end{figure}

For temperatures above the QCD phase transition, where the degrees of freedom are quarks and gluons, axions are dominantly produced by the reactions
\be
g g \leftrightarrow g a \ , \qquad \qquad 
q \overline{q} \leftrightarrow g a \ , \qquad \qquad
q g \leftrightarrow q a \ , \qquad \qquad \overline{q} g \leftrightarrow \overline{q} a \ .
\label{eq:reacagg}
\ee
These processes are model independent, since they only require QCD interactions and the $a G \tilde{G}$ coupling of (\ref{effaxion}), present in any axion model. Other processes can provide additional sources of thermal axions (see e.g. Ref.~\cite{Turner:1986tb}), but they are model dependent and we will not consider them here. The rate for the reactions in (\ref{eq:reacagg}) has been carefully computed in Ref.~\cite{Masso:2002np} and it results in\footnote{This rate was computed for the full SM spectrum. We assume the result to be valid all the way down to the QCD phase transition scale. Heavy quarks annihilating away give only a small correction.}
\be
\Gamma_a(T) = 7.1 \times 10^{-6} \, \frac{T^3}{f^2} \ , \qquad\qquad\qquad T > \Lambda_{\rm QCD} \ .
\ee
At high temperatures $g_{*s}$ is constant, and the Boltzmann equation in \Eq{eq:BoltzGlobal2} has the solution
\be
\eta(T) = 1 - \exp\left[ - \kappa \,  \left(\frac{f - T}{f} \right)\right] \ , \qquad \qquad \qquad f > T > 200 \, {\rm GeV} \ ,
\label{eq:solaboveQCD}
\ee
satisfying the correct boundary condition $\eta(f) = 0$. The dimensionless quantity $\kappa$ is defined as
\be
\kappa \equiv \frac{f}{T} \frac{\Gamma_a}{H} \simeq \frac{5 \times 10^{12} \, {\rm GeV}}{f} \ .
\ee
Thus after PQ breaking axions start to get produced, and the growth expressed in (\ref{eq:solaboveQCD}) goes on until we reach the decoupling temperature
\be
T_{\rm dec} \simeq \frac{f}{\kappa} \simeq \frac{f^2}{5 \times 10^{12} \, {\rm GeV}} \ ,
\ee
or equivalently when $\Gamma_g \simeq H$. For $\kappa = 5$ we have $\eta(T_{\rm dec}) \simeq 0.99$, therefore a thermal population is successfully generated for $f \leq 10^{12} \, {\rm GeV}$. As the temperature drops below the top mass, SM particles start annihilating away, heating the photon bath. The axion bath can be heated up, depending on whether they are still in thermal equilibrium. The validity of this condition can be check in Figure \ref{fig:Tdec1}, where we plot the decoupling temperature as a function of $f$. 

So far we have not considered the possibility of an inflationary period wiping out the produced axions. However, even if this happens, axions can be generated again below the QCD phase transition from reactions with hadrons in the initial state. We consider the scattering process within the $s$-wave approximation, and upon using (\ref{eq:axionrate}) we find the rate
\be
 \Gamma_a(i j \leftrightarrow k a) = \frac{n_i^{\rm eq} n_j^{\rm eq}}{n^{\rm eq}_a} \;
 \frac{\left|\mathcal{M}_{i j \rightarrow k a}\right|^2}{32 \pi m_i m_j} \left[1 - \left(\frac{m_k}{m_i + m_j}\right)^2 \right] \ .
\label{eq:ratebelowQCD}
\ee
The dominant contributions to the total rate comes from the processes
\be
\pi \eta \; \rightarrow  \; \pi a \ , \qquad \qquad
\pi K \; \rightarrow \; K a \ , \qquad \qquad
\pi N \; \rightarrow \; N a \ , \qquad \qquad
\pi \pi \; \rightarrow \; \pi a \ .
\ee
The matrix elements $\mathcal{M}_{i j \rightarrow k a}$ for the first three processes can be computed by using the axion couplings to hadrons as derived in Refs.~\cite{Srednicki:1985xd,Kaplan:1985dv,Georgi:1986df}. The process $\pi \pi \; \rightarrow \; \pi a$ is not present if one considers the leading order Lagrangian, so we construct the next to leading order Lagrangian in the chiral expansion and compute the associated matrix element. Once we sum over all the contributions, we find that right below the QCD phase transition $\Gamma_a \geq H$ for $f \leq 10^{12} \, {\rm GeV}$, so that axions are kept in thermal equilibrium (or in the case of a low reheating temperature axions are regenerated). These reactions are effective until $\Gamma_a \simeq H$, and we plot the decoupling temperature in Figure \ref{fig:Tdec2}. The decoupling temperature is not very sensitive to $f$, and has typical values of tens of MeV.  We have included axion-hadron interactions that arise from the $a G \tilde{G}$ coupling; the model-dependent couplings of axions to quarks could give order unity corrections to our results.

To summarize, scattering processes above and below the QCD phase transition populate the universe with thermal axions. The reactions are effective until the decoupling temperatures shown in Figure \ref{fig:Tdec2}, and after that the axion number density just red shifts with the expansion. The number density today reads
\be
n_a(T_0) = \frac{g_{*\,s}(T_0)}{g_{*\,s}(T_{\rm dec})} \frac{n_\gamma(T_0)}{2} \ .
\ee
The axion density is easily obtained as $\rho_a = m_a n_a$, and we express it here as a fraction of the critical density
\be
\Omega_a h^2 =  \frac{\rho_a}{\rho_{{\rm cr}}}  h^2 \simeq  \, 0.12 \, \frac{3.4 \times 10^5 \, \GeV}{f} = 0.12 \, \frac{m_a}{18 \, {\rm eV}} \ .
\label{eq:axiondensity}
\ee

\end{document}